\documentclass[11pt, prd, aps, showpacs, nofootinbib]{revtex4-1}

\usepackage{amsmath,amssymb}
\usepackage{graphicx}
\DeclareGraphicsExtensions{.png,.pdf}

\newcommand{\be}{\begin{equation}}
\newcommand{\ee}{\end{equation}}
\newcommand{\bea}{\begin{eqnarray}}
\newcommand{\eea}{\end{eqnarray}}

\newcommand{\Romatre}{Dipartimento di Matematica e Fisica, Universit\`a  degli Studi Roma Tre and INFN, Sezione di Roma Tre,\\ Via della Vasca Navale 84, I-00146 Rome, Italy}
\newcommand{\LaSapienza}{Dipartimento di Fisica, Universit\`a degli Studi di Roma ``La Sapienza'' and INFN, Sezione di Roma,\\ Piazzale Aldo Moro 5, 00185 Roma, Italy}
\newcommand{\RomatreINFN}{Istituto Nazionale di Fisica Nucleare, Sezione di Roma Tre,\\ Via della Vasca Navale 84, I-00146 Rome, Italy}

\begin{document}

\title{\Large Electromagnetic and strong isospin-breaking corrections\\[2mm] to the muon $g - 2$ from Lattice QCD+QED}

\author{D.~Giusti} \affiliation{\Romatre}
\author{V.~Lubicz} \affiliation{\Romatre}
\author{G.~Martinelli} \affiliation{\LaSapienza}
\author{F.~Sanfilippo} \affiliation{\RomatreINFN}
\author{S.~Simula} \affiliation{\RomatreINFN}

\begin{abstract}
\begin{center}
\vspace{0.15cm}
\includegraphics[draft=false,width=.20\linewidth]{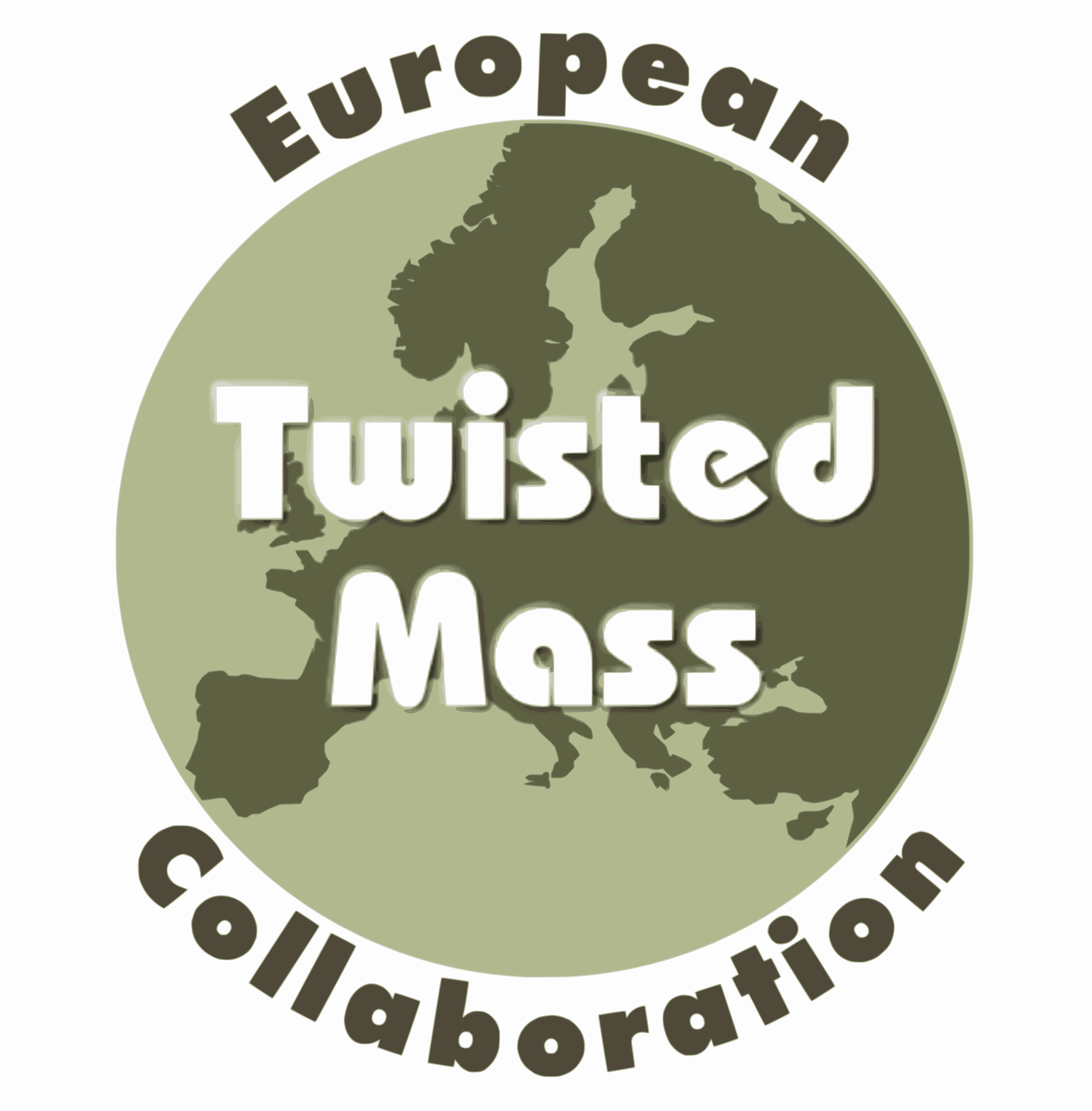}
\end{center}
We present a lattice calculation of the leading-order electromagnetic and strong isospin-breaking corrections to the hadronic vacuum polarization (HVP) contribution to the anomalous magnetic moment of the muon.
We employ the gauge configurations generated by the European Twisted Mass Collaboration (ETMC) with $N_f = 2+1+1$ dynamical quarks at three values of the lattice spacing ($a \simeq 0.062, 0.082, 0.089$ fm) with pion masses between $\simeq 210$ and $\simeq 450$ MeV.
The results are obtained adopting the RM123 approach in the quenched-QED approximation, which neglects the charges of the sea quarks. Quark disconnected diagrams are not included.
After the extrapolations to the physical pion mass and to the continuum and infinite-volume limits the contributions of the light, strange and charm quarks are respectively equal to $\delta a_\mu^{\rm HVP}(ud) = 7.1 ~ (2.5) \cdot 10^{-10}$, $\delta a_\mu^{\rm HVP}(s) = -0.0053 ~ (33) \cdot 10^{-10}$ and $\delta a_\mu^{\rm HVP}(c) = 0.0182 ~ (36) \cdot 10^{-10}$.
At leading order in $\alpha_{em}$ and $(m_d - m_u) / \Lambda_{QCD}$ we obtain $\delta a_\mu^{\rm HVP}(udsc) = 7.1 ~ (2.9) \cdot 10^{-10}$, which is currently the most accurate determination of the isospin-breaking corrections to $a_\mu^{\rm HVP}$.
\end{abstract}

\maketitle

\newpage

\section{Introduction}
\label{sec:intro}

The muon anomalous magnetic moment $a_\mu \equiv (g - 2) / 2$ is one of the most precisely determined quantities in particle physics.
It is experimentally known with an accuracy of $0.54$ ppm~\cite{Bennett:2006fi} (BNL E821) and the current precision of the Standard Model (SM) prediction is at the level of $0.4$ ppm~\cite{PDG}.
The discrepancy between the experimental value, $a_\mu^{exp}$, and the SM prediction, $a_\mu^{\rm SM}$, corresponds to $\simeq 3.5 \div 4$ standard deviations, namely $a_\mu^{exp} - a_\mu^{\rm SM} = 31.3 ~ (7.7) \cdot 10^{-10}$~\cite{Jegerlehner:2017lbd}, $a_\mu^{exp} - a_\mu^{\rm SM} = 26.8 ~ (7.6) \cdot 10^{-10}$~\cite{Davier:2017zfy} and $a_\mu^{exp} - a_\mu^{\rm SM} = 27.1 ~ (7.3) \cdot 10^{-10}$~\cite{Keshavarzi:2018mgv}.

Since the above tension may be an exciting indication of new physics (NP) beyond the SM, an intense research program is currently underway in order to achieve a significant improvement of the uncertainties.
The forthcoming $g - 2$ experiments at Fermilab (E989)~\cite{Logashenko:2015xab} and J-PARC (E34)~\cite{Otani:2015lra} aim at reducing the experimental uncertainty by a factor of four, down to 0.14 ppm, making the comparison of the experimental value $a_\mu^{exp}$ with the theoretical prediction $a_\mu^{\rm SM}$ one of the most stringent tests of the SM in the quest of NP effects.
On the theoretical side, the main uncertainty on $a_\mu^{\rm SM}$ comes from hadronic contributions, related to the hadronic vacuum polarization (HVP) and light-by-light terms~\cite{Jegerlehner:2017gek,Jegerlehner:2017lbd}.
With the planned reduction of the experimental error, the uncertainty of the hadronic corrections will soon become the main limitation of this SM test. 

The theoretical predictions for the hadronic contribution $a_\mu^{\rm HVP}$ have been traditionally obtained from experimental data using dispersion relations for relating the HVP function to the experimental cross section data for $e^+ e^-$ annihilation into hadrons~\cite{Davier:2010nc,Hagiwara:2011af}. 
An alternative approach, proposed in Refs.~\cite{Lautrup:1971jf,deRafael:1993za,Blum:2002ii}, is to compute $a_\mu^{\rm HVP}$ in Lattice QCD from the Euclidean correlation function of two electromagnetic (em) currents. 
In this respect an impressive progress in the lattice determinations of $a_\mu^{\rm HVP}$, which at leading order in $\alpha_{em}$ is a quantity of order ${\cal O}(\alpha^2_{em})$, has been achieved in the last few years~\cite{Chakraborty:2014mwa,Chakraborty:2015ugp,Blum:2015you,Blum:2016xpd,Chakraborty:2016mwy,DellaMorte:2017dyu,Boyle:2017gzv,Giusti:2017jof,Chakraborty:2017tqp,Borsanyi:2017zdw,Blum:2018mom,Giusti:2018mdh}.

With the increasing precision of the lattice calculations, it becomes necessary to include strong and em isospin-breaking (IB) corrections, which contribute to the HVP at order ${\cal O}(\alpha^2_{em} (m_d - m_u) / \Lambda_{QCD})$ and ${\cal O}(\alpha^3_{em})$, respectively.
In Ref.~\cite{Giusti:2017jof} a lattice calculation of the IB corrections to the HVP contributions due to strange and charm quarks, $\delta a_\mu^{\rm HVP}(s)$ and $\delta a_\mu^{\rm HVP}(c)$\footnote{In the strange and charm sectors the strong IB corrections are absent at leading order in ($m_d - m_u$).}, was carried out using the RM123 approach~\cite{deDivitiis:2011eh,deDivitiis:2013xla}, which is based on the expansion of the path integral in powers of the mass difference ($m_d - m_u$) and of the em coupling $\alpha_{em}$. 
The quenched QED (qQED) approximation, which neglects the effects of sea-quark charges, was adopted and quark disconnected contractions were not included because of the large statistical fluctuations of the corresponding signals.
The dominant source of uncertainty in the results of Ref.~\cite{Giusti:2017jof} was related to the em corrections to the renormalization constant (RC) of the local vector current, computed through the axial Ward-Takahashi identity derived in the QCD+QED theory.

In this work we present our determination of the IB corrections to the HVP contribution due to the light $u$- and $d$-quarks, $\delta a_\mu^{\rm HVP}(ud)$, using the same methods and lattice setup adopted in Ref.~\cite{Giusti:2017jof} in the case of the strange and charm contributions.
A preliminary result for $\delta a_\mu^{\rm HVP}(ud)$ was presented in Ref.~\cite{Giusti:2018vrc}.
Thanks to a recent nonperturbative evaluation of QCD+QED effects on the RCs of bilinear operators performed in Ref.~\cite{DiCarlo:2019thl} we can update the determinations of $\delta a_\mu^{\rm HVP}(s)$ and $\delta a_\mu^{\rm HVP}(c)$ made in Ref.~\cite{Giusti:2017jof}, obtaining a drastic improvement of the uncertainty by a factor of $\approx 3$ and $\approx 3.5$, respectively.

Within the qQED approximation and neglecting quark-disconnected diagrams the main results of the present study are:
\bea
    \label{eq:delta_ud_intro}
    \delta a_\mu^{\rm HVP}(ud) & = & 7.1 ~ (1.1)_{stat+fit}\, (1.3)_{input}\,(1.2)_{chir}\, (1.2)_{\rm FVE}\, (0.6)_{a^2} \cdot 10^{-10} \nonumber \\
                                                 & = & 7.1 ~ (2.5) \cdot 10^{-10} ~ , \\[2mm]
    \label{eq:delta_s_intro}
    \delta a_\mu^{\rm HVP}(s) & = & -0.0053 ~ (30)_{stat+fit}\, (13)_{input}\, (2)_{chir}\, (2)_{\rm FVE}\, (1)_{a^2} \cdot 10^{-10} \nonumber \\
                                               & = & -0.0053 ~ (33) \cdot 10^{-10} ~ , \\[2mm]
    \label{eq:delta_c_intro}
    \delta a_\mu^{\rm HVP}(c) & = & 0.0182 ~ (35)_{stat+fit}\, (5)_{input}\,(1)_{chir}\, (3)_{\rm FVE}\, (1)_{a^2} \cdot 10^{-10} \nonumber \\
                                               & = & 0.0182 ~ (36) \cdot 10^{-10} ~ , 
\eea
where the errors come from (statistics + fitting procedure), input parameters, chiral extrapolation, finite-volume and discretization effects.
Thus, we confirm that the em corrections $\delta a_\mu^{\rm HVP} (s)$ and $\delta a_\mu^{\rm HVP} (c)$ turn out to be negligible with respect to the current uncertainties of the corresponding lowest-order terms $a_\mu^{\rm HVP}(s) = 53.1 ~ (2.5) \cdot 10^{-10}$ and $a_\mu^{\rm HVP}(c) = 14.75 ~ (0.56) \cdot 10^{-10}$ determined in Ref.~\cite{Giusti:2017jof}. 
In the case of the $u$- and $d$-quarks our finding (\ref{eq:delta_ud_intro}) corresponds to about $1.2 \%$ of the lowest-order value $a_\mu^{\rm HVP}(ud) = 619 ~ (17.8) \cdot 10^{-10}$ obtained recently in Ref.~\cite{Giusti:2018mdh}.

Recent calculations of the IB corrections to the HVP are: $\delta a_\mu^{\rm HVP}(ud) = 9.0~(4.5) \cdot 10^{-10}$ from FNAL/HPQCD/MILC~\cite{Chakraborty:2017tqp}, which includes only strong IB effects, and $\delta a_\mu^{\rm HVP}(ud) = 9.5~(10.2) \cdot 10^{-10}$ from RBC/UKQCD~\cite{Blum:2018mom}, which includes also one disconnected QED diagram.
In Ref.~\cite{Borsanyi:2017zdw} the BMW collaboration has estimated the value $\delta a_\mu^{\rm HVP}(ud) = 7.8~(5.1) \cdot 10^{-10}$ from results of the dispersive analysis of $e^+ e^-$ data~\cite{Jegerlehner:2017lbd}.
In the case of the strange contribution $\delta a_\mu^{\rm HVP}(s)$ RBC/UKQCD has recently obtained the result $\delta a_\mu^{\rm HVP} (s) = -0.0149~(32) \cdot 10^{-10}$~\cite{Blum:2018mom}, which confirms the smallness of such contribution though it differs slightly from our finding (\ref{eq:delta_s_intro}).

Summing up the three contributions (\ref{eq:delta_ud_intro})-(\ref{eq:delta_c_intro}) and adding a further $\approx 15 \%$ uncertainty related to the qQED approximation and to the neglect of quark-disconnected diagrams (see Section~\ref{sec:results}), we get
\be
    \label{eq:delta_udsc_intro}
    \delta a_\mu^{\rm HVP}(udsc) = 7.1 ~ (2.6) ~ (1.2)_{qQED+disc} \cdot 10^{-10} = 7.1 ~ (2.9) \cdot 10^{-10} ~ ,
\ee
which represents the most accurate determination of the IB corrections to $a_\mu^{\rm HVP}$ to date.

The paper is organized as follows.
In section~\ref{sec:IB} we describe the evaluation  of the em and strong IB corrections to the light-quark HVP contribution at order ${\cal O} (\alpha^2_{em} (m_d - m_u) / \Lambda_{QCD})$ and ${\cal O} (\alpha^3_{em})$ using the RM123 approach~\cite{deDivitiis:2011eh,deDivitiis:2013xla}. 
Details of the lattice simulations are collected in the Appendix~\ref{sec:appA}.
In section~\ref{sec:results} we describe the extrapolation to the physical pion mass and to the continuum and infinite volume limits.
Finally, section~\ref{sec:conclusions} contains our conclusions and outlooks for future developments.

\section{Isospin-breaking corrections in the RM123 approach}
\label{sec:IB}

We adopt the time-momentum representation for the evaluation of the HVP contribution $a^{\rm HVP}_\mu$ to the muon ($g - 2$), namely~\cite{Bernecker:2011gh} 
\be
    a^{\rm HVP}_\mu = 4 \alpha_{em}^2 \int_0^\infty dt ~ K_\mu(t) V(t) ~ ,
    \label{eq:amu_t}
\ee
where the kernel function $K_\mu(t)$ is given by
\be
    K_\mu(t) = \frac{4}{m_\mu^2} \int_0^\infty d\omega ~ \frac{1}{\sqrt{4 + \omega^2}} ~ 
                      \left( \frac{\sqrt{4 + \omega^2} - \omega}{\sqrt{4 + \omega^2} + \omega} \right)^2 
                      \left[ \frac{\mbox{cos}(\omega m_\mu t) - 1} {\omega^2} + \frac{1}{2} m_\mu^2 t^2 \right] 
    \label{eq:ftilde}
\ee
with $m_\mu$ being the muon mass.
In Eq.~(\ref{eq:amu_t}) the quantity $V(t)$ is the vector current-current Euclidean correlator defined as
\be
    V(t) \equiv - \frac{1}{3} \sum_{i=1,2,3} \int d\vec{x} ~ \langle J_i(\vec{x}, t) J_i(0) \rangle ~ ,
    \label{eq:VV}
\ee
where
 \be
     J_\mu(x) \equiv \sum_{f = u, d, s, c, ...} J_\mu^f(x) =  \sum_{f = u, d, s, c, ...} q_f ~ \overline{\psi}_f(x) \gamma_\mu \psi_f(x)
     \label{eq:Jem}
 \ee
is the em current operator with $q_f$ being the electric charge of the quark with flavor $f$ in units of the electron charge $e$, while $\langle ... \rangle$ means the average of the $T$-product over gluon and quark fields.

We will limit ourselves to the HVP contribution of the light $u$ and $d$ quarks, indicated by $a^{\rm HVP}_\mu (ud)$, neglecting off-diagonal flavor terms (i.e.~including quark-connected diagrams only).
In this case each quark flavor $f$ contributes separately
\be
    a^{\rm HVP}_\mu (ud) = \sum_{f = u, d} [a^{\rm HVP}_\mu(f)]_{(conn)} ~ .
    \label{eq:amuf}
\ee
For sake of simplicity we drop the suffix $(conn)$, but it is understood that in the following we refer always to quark-connected contractions only.

In the RM123 method of Refs.~\cite{deDivitiis:2011eh,deDivitiis:2013xla} the vector correlator $V(t)$ is expanded into a lowest-order contribution $V^{ud}(t)$, evaluated in isospin symmetric QCD (i.e.~$m_u = m_d$ and $\alpha_{em}=0$), and a correction $\delta V^{ud}(t)$ computed at leading order in the small parameters $(m_d - m_u) / \Lambda_{QCD}$ and $\alpha_{em}$:
\be
     V(t) = V^{ud}(t) + \delta V^{ud}(t) + \dots ~ ,
\ee
where the ellipses stand for higher order terms in $(m_d - m_u) / \Lambda_{QCD}$ and $\alpha_{em}$. 

The separation between the isosymmetric QCD and the IB contributions, $V^{ud}(t)$ and $\delta V^{ud}(t)$, is prescription dependent. 
In this work we follow Ref.~\cite{Giusti:2017jof} and we impose the matching condition in which the renormalized coupling and quark masses in the full theory, $\alpha_s$ and $m_f$, and in isosymmetric QCD, $\alpha_s^{(0)}$ and $m_f^{(0)}$, coincide in the $\overline{\rm MS}$ scheme at a scale of $2~\mbox{GeV}$.
Such a prescription is known as the GRS one~\cite{Gasser:2003hk}.

\begin{figure}[htb!]
\begin{center}
\includegraphics[scale=1.8]{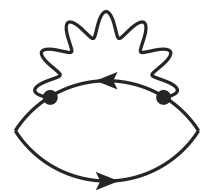} ~ 
\includegraphics[scale=1.8]{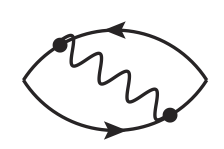} ~ 
\includegraphics[scale=1.8]{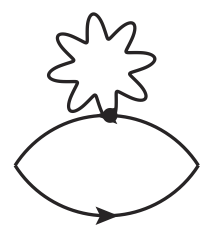} ~ 
\includegraphics[scale=1.8]{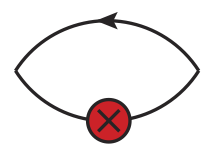} ~ 
\includegraphics[scale=1.8]{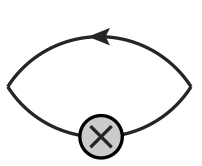}\\
~ (a) \qquad \qquad \qquad (b) \qquad \qquad \qquad ~ (c) \qquad \qquad \qquad ~ (d) \qquad \qquad \qquad (e) ~
\end{center}
\vspace{-0.75cm}
\caption{\it \small Fermionic connected diagrams contributing to the IB corrections to $a^{\rm HVP}_\mu(ud)$: self-energy (a), exchange (b), tadpole (c), pseudoscalar (d) and scalar (e) insertions. Solid lines represent light-quark propagators in isosymmetric QCD.}
\label{fig:diagrams}
\end{figure}

The calculation of the IB correlator $\delta V^{ud}(t)$ requires the evaluation of the self-energy, exchange, tadpole, pseudoscalar and scalar insertion diagrams depicted in Fig.~\ref{fig:diagrams}.

More specifically one has
\be
    \delta V^{ud} (t) \equiv \delta V^J(t) + \delta V^T(t) + \delta V^{PS}(t) + \delta V^S(t) + \delta V^{SIB}(t) + \delta V^{{\cal{Z}}_A}(t) ~ 
\label{eq:deltaV}
\ee
where $\delta V^{{\cal{Z}}_A}(t)$ will be described later in this Section and
\bea
    \label{eq:deltaV_J}
    \delta V^J(t) & = & \frac{4 \pi \alpha_{em}}{3} \sum_{f = u, d} \, \sum_{i = 1, 2, 3} \frac{1}{2} \sum_{\vec{x}, y_1, y_2}
                                  \langle 0| T \left \{ \left[ J_i^f(\vec{x}, t) \right]^\dagger {\cal J}_\mu^f(y_1) {\cal J}_\nu^f(y_2) ~ 
                                  J_i^f(0) \right \} | 0 \rangle \Delta_{\mu \nu}(y_1, y_2) ~ , \qquad \\
    \label{eq:deltaV_T}
    \delta V^T(t) & = & \frac{4 \pi \alpha_{em}}{3} \sum_{f = u, d} \, \sum_{i = 1, 2, 3} ~ \sum_{\vec{x}, y} ~ 
                                  \langle 0| T \left \{ \left[ J_i^f(\vec{x}, t) \right]^\dagger {\cal T}_\nu^f(y) ~ J_i^f(0) \right \} | 0 \rangle 
                                  \Delta_{\nu \nu}(y, y) ~ , \\
    \label{eq:deltaV_PS}
    \delta V^{PS}(t) & = & \frac{4 \pi \alpha_{em}}{3} \sum_{f = u, d} ~ \delta m_f^{crit} \sum_{i = 1, 2, 3} ~ \sum_{\vec{x}, y} ~ 
                                        \langle 0| T \left \{ \left[ J_i^f(\vec{x}, t) \right]^\dagger \, i \, \overline{\psi}_f(y) \gamma_5  \psi_f(y) ~ 
                                        J_i^f(0) \right \} | 0 \rangle ~ , \\
    \label{eq:deltaV_S}
    \delta V^S(t) & = & - \frac{4 \pi \alpha_{em}}{3} \sum_{f = u, d} \frac{{\cal{Z}}_m}{Z_m} \, m_f^{(0)} \sum_{i = 1, 2, 3} ~ \sum_{\vec{x}, y} ~
                                   \langle 0| T \left \{ \left[ J_i^f(\vec{x}, t) \right]^\dagger ~ \overline{\psi}_f(y)  \psi_f(y) ~ J_i^f(0) \right \} | 0 \rangle ~ , \\
      \label{eq:deltaV_SIB}
      \delta V^{SIB}(t) & = & - \frac{1}{3} \sum_{f = u, d} \frac{1}{Z_m} \left[ m_f - m_f^{(0)} \right] \sum_{i = 1, 2, 3} \sum_{\vec{x}, y} ~ 
                                          \langle 0| T \left \{ \left[ J_i^f(\vec{x}, t) \right]^\dagger  \overline{\psi}_f(y) \psi_f(y) J_i^f(0) \right \} | 0 \rangle ~~~~
\eea
with ${\cal J}_\mu^f(y)$ and ${\cal T}_\nu^f(y)$ being the lattice conserved current and the tadpole operator for the quark flavor $f$, respectively\footnote{In Eqs.~(\ref{eq:conservedV}-\ref{eq:tadpole}) the matrix $i \tau^3 \gamma_5$ appears because in the twisted-mass action the Wilson term is {\em twisted} (in the so-called physical basis at maximal twist). In the case of standard Wilson fermions the matrix $i \tau^3 \gamma_5$ should be replaced by the unit one.},
\bea
    \label{eq:conservedV}
    {\cal J}_\mu^f(y) & = & q_f ~ \frac{i}{2} \left[ \overline{\psi}_f(y) (i \tau^3 \gamma_5 - \gamma_\mu) U_\mu(y) \psi_f(y + a \hat{\mu}) -
                                         \overline{\psi}_f(y + a \hat{\mu}) (i \tau^3 \gamma_5 + \gamma_\mu) U_\mu^\dagger(y) \psi_f(y) \right] ~ , \quad \\[2mm]
    \label{eq:tadpole}
     {\cal T}_\nu^f(y) & = & q_f^2 ~ \frac{1}{2} \left[ \overline{\psi}_f(y) (i \tau^3 \gamma_5 - \gamma_\nu) U_\nu(y) \psi_f(y+ a \hat{\nu}) +
                                         \overline{\psi}_f(y + a \hat{\nu}) (i \tau^3 \gamma_5 + \gamma_\nu) U_\nu^\dagger(y) \psi_f(y) \right] ~ ,
\eea
while $\Delta_{\mu \nu}(y_1, y_2)$ is the photon propagator.
In Eq.~(\ref{eq:deltaV_PS}) $\delta m_f^{crit}$ is the em shift of the critical mass for the quark flavor $f$.
In Eq.~(\ref{eq:deltaV_S}) the quantity ${\cal{Z}}_m$ is related to the em corrections to the mass RC in QCD+QED, $Z_m^{QCD+QED}$, as
\be
    Z_m^{QCD+QED} = Z_m \left( 1 - 4 \pi \alpha_{em} {\cal{Z}}_m \right) + {\cal{O}}(\alpha_{em}^m \alpha_s^n) ~ , \qquad (m > 1, ~ n \geq 0)
    \label{eq:Zmem}
\ee
where $Z_m$ is the mass RC in QCD only and the product $Z_m \, {\cal{Z}}_m$ encodes the corrections at first order in $\alpha_{em}$. 
The quantity ${\cal{Z}}_m$ can be written as
\be
    {\cal Z}_m = {\cal Z}_m^{(1)} \cdot Z_m^{fact} ~ ,
    \label{eq:Zm}
\ee
where ${\cal{Z}}_m^{(1)}$ is the pure QED contribution at leading order in $\alpha_{em}$, given in the $\overline{\rm MS}$ scheme at the renormalization scale $\mu$ by~\cite{Martinelli:1982mw,Aoki:1998ar}
\be
    {\cal{Z}}_m^{(1)} = \frac{q_f^2}{16 \pi^2} \left[ 6 ~ \mbox{ln}(a \mu) - 22.5954 \right] ~ ,
    \label{eq:Zm_1}
\ee
while $Z_m^{fact}$ accounts for the corrections of order ${\cal{O}}(\alpha_s^n)$ with $n \geq 1$ to Eq.~(\ref{eq:Zm}).
It represents the QCD correction to the ``naive factorization'' approximation ${\cal{Z}}_m = {\cal{Z}}_m^{(1)}$ (i.e.~$Z_m^{fact} = 1$) adopted in Ref.~\cite{Giusti:2017jof}.
Finally, Eq.~(\ref{eq:deltaV_SIB}) corresponds to the strong IB (SIB) effect (in the GRS prescription) with $m_u^{(0)} = m_d^{(0)} = m_{ud}^{(0)}$ being the renormalized light-quark mass in isosymmetric QCD.

In the numerical evaluation of the photon propagator, performed in the Feynman gauge, the photon zero-mode has been removed according to the QED$_L$ prescription~\cite{Hayakawa:2008an}, i.e.~the photon field $A_\mu$ satisfies $A_\mu(k_0, \vec{k} = \vec{0}) \equiv 0$ for all $k_0$.

In this work we make use of the same isosymmetric QCD gauge ensembles used in Ref.~\cite{Giusti:2017jof}, i.e.~those generated by the European Twisted Mass Collaboration (ETMC) with $N_f = 2+1+1$ dynamical quarks, which include in the sea, besides two light mass-degenerate quarks, also the strange and the charm quarks with masses close to their physical values~\cite{Baron:2010bv,Baron:2011sf}.
For earlier investigations of finite volume effects (FVEs) the ETMC produced three dedicated ensembles, A40.20, A40.24 and A40.32 (see Appendix~\ref{sec:appA} for details), which share the same light-quark mass and lattice spacing and differ only in the lattice size $L$.
To improve such an investigation a further gauge ensemble, A40.40, has been generated at a larger value of the lattice size $L$.

For our maximally twisted-mass setup $\delta m_f^{crit}$ has been determined in Ref.~\cite{Giusti:2017dmp}, while $1 / Z_m = Z_P$, where $Z_P$ is the RC of the pseudoscalar density evaluated in Ref.~\cite{Carrasco:2014cwa}.
The coefficient $Z_m^{fact}$ has been recently computed in Ref.~\cite{DiCarlo:2019thl} in a non-perturbative framework within the RI$^\prime$-MOM scheme~\cite{Martinelli:1994ty}. 

Within the qQED approximation, which treats the dynamical quarks as electrically neutral particles, the correlator $\delta V^J(t)$ corresponds to the sum of the diagrams (\ref{fig:diagrams}a)-(\ref{fig:diagrams}b), while the correlators $\delta V^T(t)$ and $\delta V^{PS}(t)$ represent the contributions of the diagrams (\ref{fig:diagrams}c) and (\ref{fig:diagrams}d), respectively.
The diagram (\ref{fig:diagrams}e) contributes to both $\delta V^S(t)$ and $\delta V^{SIB}(t)$.

In our numerical simulations we have adopted the following local version of the vector current:
\be
    J_\mu (x) = Z_A ~ q_f ~ \overline{\psi}_{f^\prime} (x) \gamma_\mu \psi_f (x) ~ ,
    \label{eq:localV}
\ee
where $\overline{\psi}_{f^\prime}$ and $\psi_f$ represent two quarks with the same mass, charge and flavor, but regularized with opposite values of the Wilson $r$-parameter (i.e.~$r_{f^\prime} = - r_f$).
Being at maximal twist the current~(\ref{eq:localV}) renormalizes multiplicatively with the RC $Z_A$ of the axial current.
By construction the local current~(\ref{eq:localV}) does not generate quark-disconnected diagrams.

As discussed in Ref.~\cite{Giusti:2017jof}, the properties of the kernel function $K_\mu(t)$, given by Eq.~(\ref{eq:ftilde}), guarantee that the contact terms, generated in the HVP tensor by a local vector current, do not contribute to both $a_\mu^{HVP}$ and its IB correction.

Since we have adopted the renormalized vector current (\ref{eq:localV}), the contribution $\delta V^{{\cal{Z}}_A}(t)$, appearing in Eq.~(\ref{eq:deltaV}), takes into account  the em corrections to the RC $Z_A$ in QCD+QED, namely
\be
    Z_A^{QCD+QED} = Z_A \, \left( 1 + 4 \pi \alpha_{em} {\cal Z}_A \right) + {\cal{O}}(\alpha_{em}^m \alpha_s^n) ~ , \qquad (m > 1, ~ n \geq 0)
    \label{eq:ZAem}
\ee
where $Z_A$ is the RC of the axial current in pure QCD (determined in Ref.~\cite{Carrasco:2014cwa}), while the product $Z_A \, {\cal{Z}}_A$ encodes  the corrections at first order in $\alpha_{em}$. 
The quantity ${\cal{Z}}_A$ can be written as
\be
    {\cal Z}_A = {\cal Z}_A^{(1)} \cdot Z_A^{fact} ~ ,
    \label{eq:ZA}
\ee
where ${\cal Z}_A^{(1)}$ is the pure QED correction at leading order in $\alpha_{em}$, given by~\cite{Martinelli:1982mw,Aoki:1998ar}
\be
    {\cal{Z}}_A^{(1)} = - 15.7963 ~ \frac{q_f^2}{16 \pi^2} ~ ,
    \label{eq:ZA_1}
\ee
and $Z_A^{fact}$ takes into account QCD corrections of order ${\cal{O}}(\alpha_s^n)$ with $n \geq 1$ to Eq.~(\ref{eq:ZA}).
In this work we make use of the non-perturbative determination obtained in Ref.~\cite{DiCarlo:2019thl} within the RI$^\prime$-MOM scheme, which improves significantly the value $Z_A^{fact} = 0.9\,(1)$ obtained through the axial Ward-Takahashi identity in Ref.~\cite{Giusti:2017jof}. 
The values adopted for the coefficients $Z_m^{fact}$ and $Z_A^{fact}$ are collected in Table~\ref{tab:Zfact} of Appendix~\ref{sec:appA}.

Thus, the IB term $\delta V^{{\cal Z}_A}(t)$ is simply given by
\be
    \delta V^{{\cal{Z}}_A}(t) \equiv - 0.2001 \, 4 \pi \alpha_{em} \, (q^4 / q^2) \, Z_A^{fact} \, V^{ud}(t) ~ ,
    \label{eq:deltaV_ZA}
\ee
where $q^4 / q^2 \equiv \sum_{f = u, d} \, q_f^4 / \sum_{f = u, d} \, q_f^2 = 17 / 45$ and $V^{ud}(t)$ is the lowest-order contribution of the light-quarks to the vector correlator, calculated for our lattice setup in Ref.~\cite{Giusti:2018mdh}.

To sum up, the IB corrections $\delta V^{ud}(t)$ can be written as the sum of two (prescription dependent) contributions as
 \be
      \delta V^{ud}(t) = \delta V^{QED}(t) + \delta V^{SIB}(t) ~ ,
      \label{eq:QED+QCD}
 \ee
where
 \be
     \delta V^{QED}(t) = \delta V^J(t) + \delta V^T(t) + \delta V^{PS}(t) + \delta V^S(t) + \delta V^{{\cal{Z}}_A}(t)
     \label{eq:deltaV_QED}
 \ee
and $\delta V^{SIB}(t)$ is given by Eq.~(\ref{eq:deltaV_SIB}).

Within the qQED approximation, where the shift $\delta m_f^{crit}$ is proportional to $q_f^2$~\cite{Giusti:2017dmp}, and neglecting quark-disconnected diagrams the QED correlator $\delta V^{QED}(t)$ is proportional to $q^4 \equiv \sum_{f = u, d} \, q_f^4 = 17 / 81$. 
Instead, the SIB correlator $\delta V^{SIB}(t)$ is proportional to $\sum_{f = u, d} \, q_f^2 \, (m_f^{(0)}- m_f) = (1/6) \, (m_d - m_u)$.
Using as inputs the experimental charged- and neutral-kaon masses the value $m_d - m_u = 2.38 \, (18)$ MeV was determined in Ref.~\cite{Giusti:2017dmp} at the physical point in the $\overline{\rm MS}(2 ~ \rm GeV)$ scheme.
Such a value is adopted in Eq.~(\ref{eq:deltaV_SIB}) for all gauge ensembles.

\begin{figure}[htb!]
\begin{center}
\includegraphics[scale=0.375]{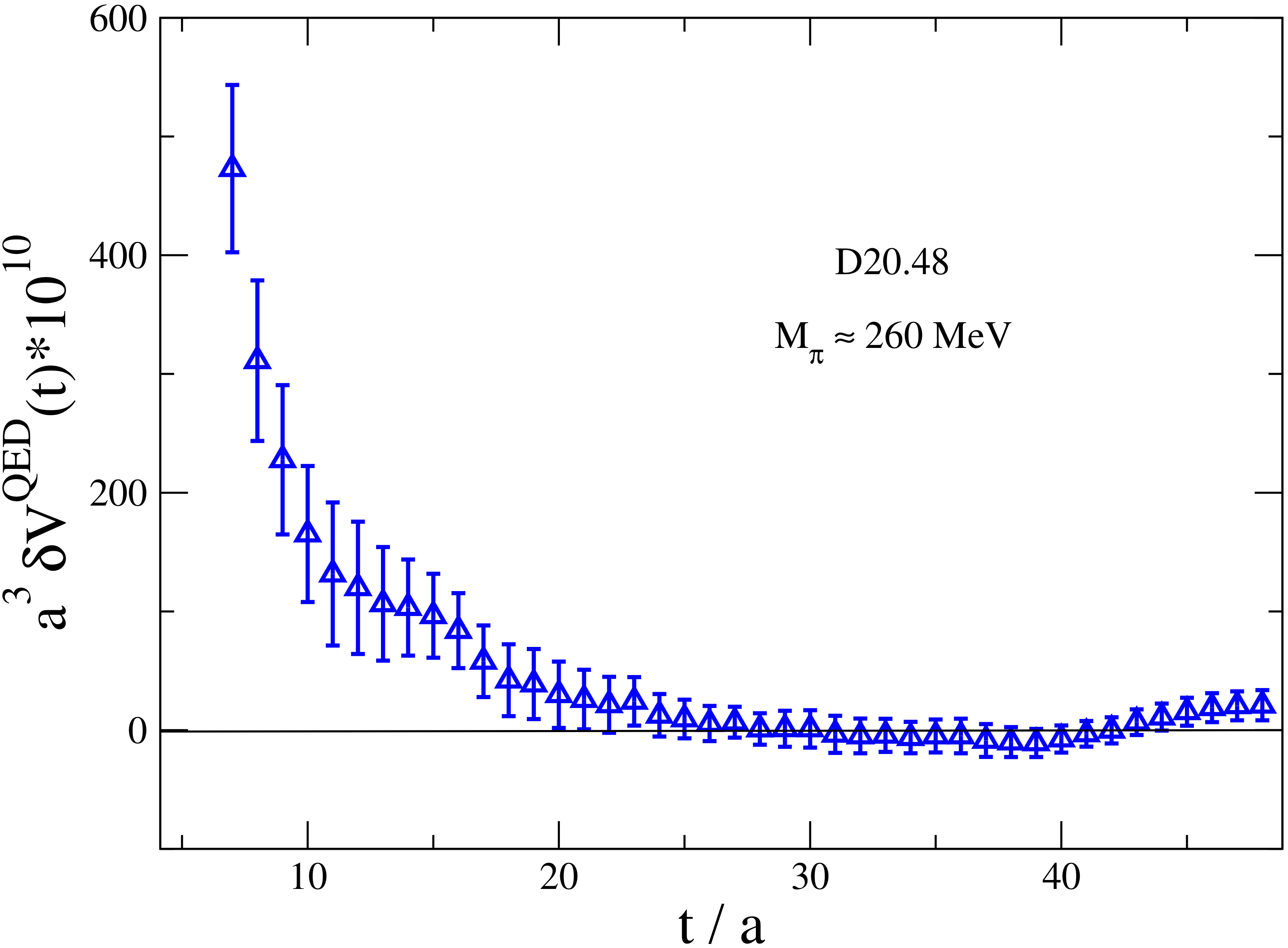} ~~ \includegraphics[scale=0.375]{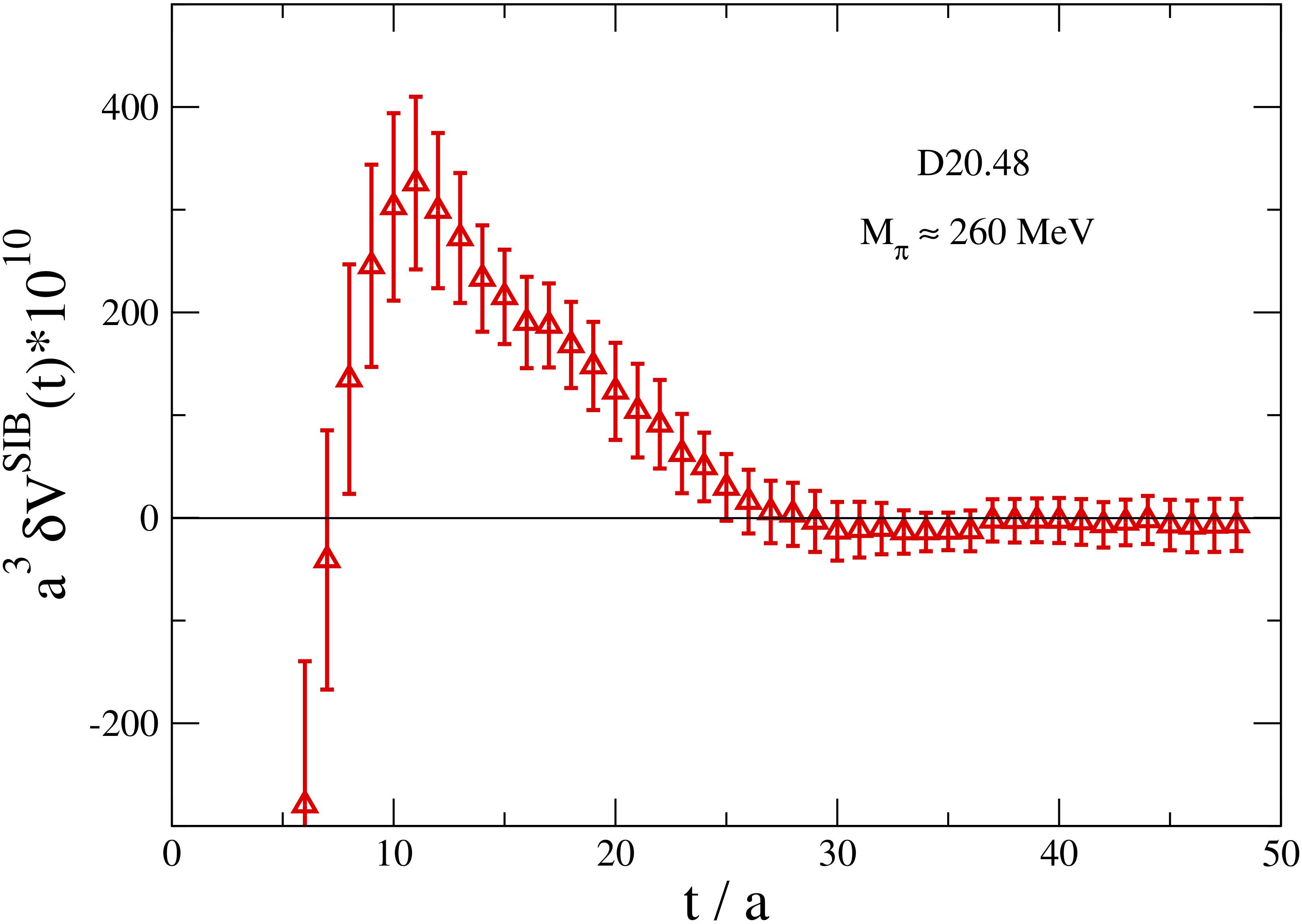}
\end{center}
\vspace{-0.50cm}
\caption{\it \small Left panel: time dependence of the IB contribution $\delta V^{QED}(t)$ [see Eq.~(\ref{eq:deltaV_QED})] in lattice units in the case of the gauge ensemble $D20.48$ (see Appendix~\ref{sec:appA}). Right panel: the same as in the left panel, but for the SIB term $\delta V^{SIB}(t)$ [see Eq.~(\ref{eq:deltaV_SIB})]. The simulated pion mass is $M_\pi \simeq 260$ MeV and the lattice spacing is equal to $a \simeq 0.06$ fm.}
\label{fig:dVud}
\end{figure}

In Fig.~\ref{fig:dVud} we show the dependence of both $\delta V^{QED}(t)$ and $\delta V^{SIB}(t)$ on the time distance $t$ in the case of the ETMC gauge ensemble $D20.48$ (see Appendix~\ref{sec:appA}).

\section{Results}
\label{sec:results}

A convenient procedure~\cite{DellaMorte:2017dyu,Giusti:2017jof,Giusti:2018mdh} consists in splitting Eq.~(\ref{eq:amu_t}) into two contributions corresponding to $0 \leq t \leq T_{data}$ and $t > T_{data}$, respectively.
In the first contribution the vector correlator is numerically evaluated on the lattice, while for the second contribution an analytic representation is required.
If $T_{data}$ is large enough that the ground-state contribution is dominant for $t > T_{data}$ and smaller than $T / 2$ in order to avoid backward signals, the IB corrections $\delta a_\mu^{\rm HVP}(ud)$ can be written as
\be
    \delta a_\mu^{\rm HVP} (ud) \equiv \delta a_\mu^{\rm HVP}(<) + \delta a_\mu^{\rm HVP}(>) 
    \label{eq:deltamu}
\ee
with
\bea 
    \label{eq:deltamu_cuts_I}
    \delta a_\mu^{\rm HVP}(<) & = & 4 \alpha_{em}^2 \sum_{t = 0}^{T_{data}} K_\mu(t) ~ \delta V^{ud}(t) ~ , \\[2mm]
    \label{eq:deltamu_cuts_II}
    \delta a_\mu^{\rm HVP}(>) & = & 4 \alpha_{em}^2 \sum_{t = T_{data} + a}^\infty K_\mu(t) ~ 
                                                         \delta \left[ \frac{Z^{ud}_V} {2 M^{ud}_V} e^{- M^{ud}_V t} \right]  \nonumber \\
                                               & = & 4 \alpha_{em}^2 \sum_{t = T_{data} + a}^\infty K_\mu(t) ~ 
                                                          \frac{Z^{ud}_V} {2 M^{ud}_V} e^{- M^{ud}_V t} \left[ \frac{\delta Z^{ud}_V}{Z^{ud}_V}
                                                         - \frac{\delta M^{ud}_V}{M^{ud}_V} (1 + M^{ud}_V t) \right]  ~ , ~                                                   
\eea
where $M_V^{ud}$ is the ground-state mass of the lowest-order correlator $V^{ud}(t)$ and $Z_V^{ud}$ is the squared matrix element of the vector current between the ground-state $| V \rangle$ and the vacuum: $Z_V^{ud} \equiv (1/3) \sum_{i=x,y,z}$ $\sum_{f=u,d} q_f^2$ $| \langle 0 | \overline{\psi}_f(0) \gamma_i \psi_f(0) | V \rangle |^2$.
In Ref.~\cite{Giusti:2018mdh} the ground-state masses $M_V^{ud}$ and the matrix elements $Z_V^{ud}$ have been determined from a single exponential fit of $V^{ud}(t)$ using appropriate time intervals $t_{min} \leq t \leq t_{max}$, where the ground-state is dominating. 
For the reader's convenience the values chosen in Ref.~\cite{Giusti:2018mdh} for $t_{min}$ and $t_{max}$ at each value of $\beta$ and of the lattice volume are shown in Table \ref{tab:GS_light}.
\begin{table}[hbt!]
\begin{center}
\begin{tabular}{||c|c||c|c||}
\hline
$\beta$ & $V / a^4$ & $t_{min} / a$ & $t_{max} / a$ \\
\hline \hline
$1.90$ & $40^3 \times 80$ &$12$ &$22$ \\
\cline{2-4}
            & $32^3 \times 64$ &$12$ &$22$ \\
\cline{2-4}
            & $24^3 \times 48$ &$12$ &$20$ \\
\cline{2-4}
            & $20^3 \times 48$ &$12$ &$20$ \\
\hline \hline
$1.95$ & $32^3 \times 64$ &$13$ &$22$ \\
\cline{2-4}
            & $24^3 \times 48$ & $13$ &$20$ \\
\hline \hline
$2.10$ & $48^3 \times 96$ & $18$ &$30$ \\ 
\hline   
\end{tabular}
\end{center}
\vspace{-0.25cm}
\caption{\it \small Values of $t_{min}$ and $t_{max}$ adopted in Ref.~\cite{Giusti:2018mdh} to extract the ground-state signal from the light-quark vector correlator $V^{ud}(t)$ for each value of $\beta$ and of the lattice volume $V / a^4$ for the ETMC gauge ensembles adopted in this work (see Table~\ref{tab:simudetails} of Appendix~\ref{sec:appA}).}
\label{tab:GS_light}
\end{table}

In Ref.~\cite{Giusti:2018mdh} for each ETMC gauge ensemble the lowest-order correlator $V^{ud}(t)$ was fitted at large time distances using also the two-pion finite volume spectrum.
It turned out that the first two-pion energy level $E_{n=1}^{\pi \pi}$ is always close to $M_V^{ud}$ within the uncertainties.
This is reassuring that the use of a single exponential fit in Eq.~(\ref{eq:deltamu_cuts_II}) reproduces properly the tail of the correlator beyond $T_{data}$. 
To illustrate this point we have collected in Table~\ref{tab:two-pion} the values of $M_V^{ud}$ and $E_{n=1}^{\pi \pi}$ for the three ensembles A30.32, B25.32 and D20.48, corresponding to quite similar values of the pion mass ($M_\pi \approx 265$ MeV) and to values of $M_\pi L$ ranging from $3.0$ to $3.9$ (see Table~\ref{tab:simudetails}).
\begin{table}[hbt!]
\begin{center}
\begin{tabular}{||c|c||c|c||}
\hline
~ ensemble ~ & ~ $M_\pi L$ ~ & ~ $M_V^{ud}$ ~ (MeV) ~ & ~ $E_{n=1}^{\pi \pi}$ ~ (MeV) ~ \\
\hline
A30.32 & 3.9 & 843 ~ (26) & 846 ~ (31) \\
\hline
B25.32 & 3.4 & 868 ~ (25) & 848 ~ (29) \\
\hline
D20.48 & 3.0 & 877 ~ (26) & 839 ~ (23) \\
\hline   
\end{tabular}
\end{center}
\vspace{-0.25cm}
\caption{\it \small Values of $M_V^{ud}$ and $E_{n=1}^{\pi \pi}$ (see text) obtained in Ref.~\cite{Giusti:2018mdh} for the three ensembles A30.32, B25.32 and D20.48, corresponding  to quite similar values of the pion mass ($M_\pi \approx 265~\rm{MeV}$) and to different values of $M_\pi L$ (see Table~\ref{tab:simudetails} of Appendix~\ref{sec:appA}).}
\label{tab:two-pion}
\end{table}

\begin{figure}[htb!]
\begin{center}
\includegraphics[scale=0.375]{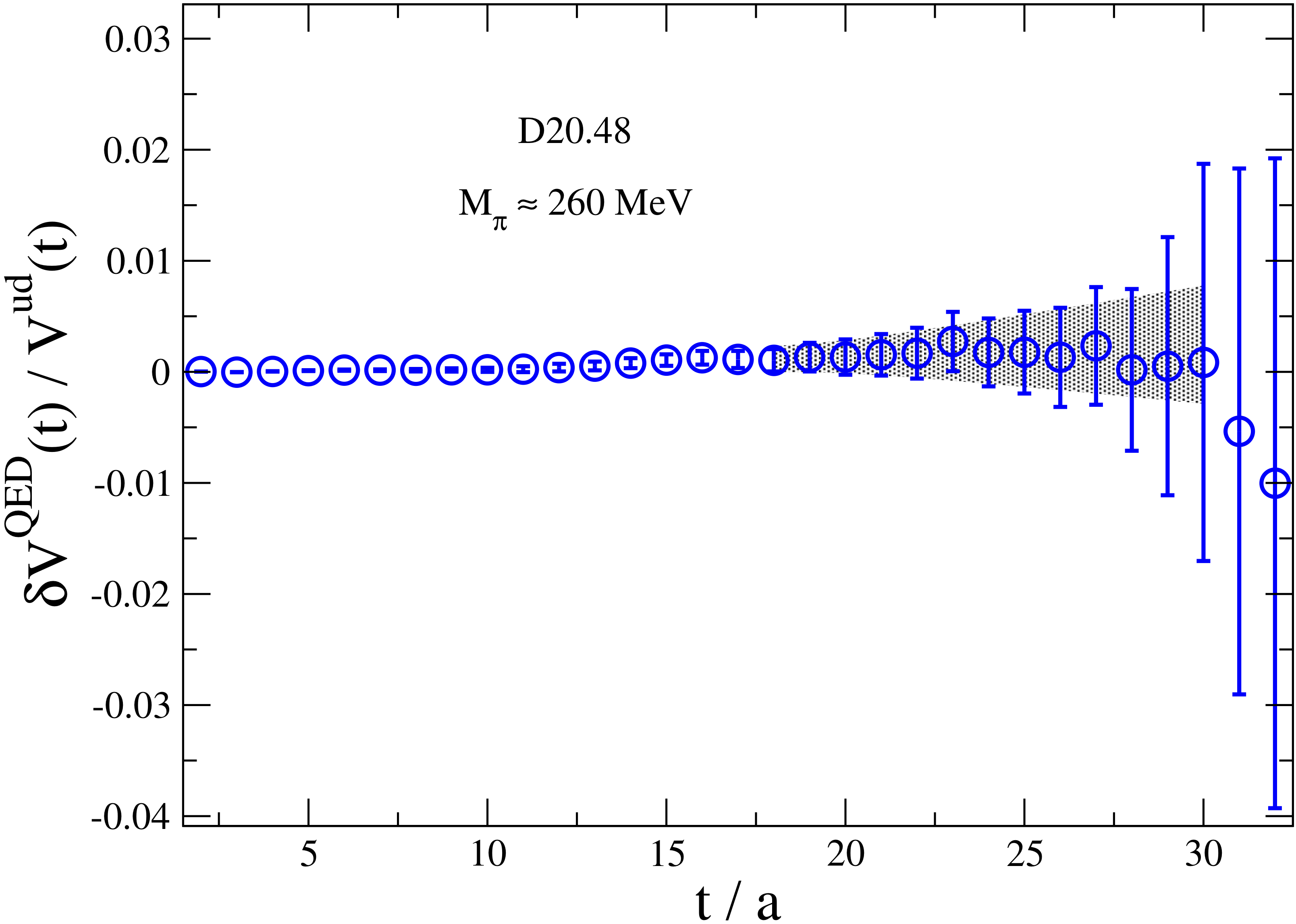} ~~ \includegraphics[scale=0.375]{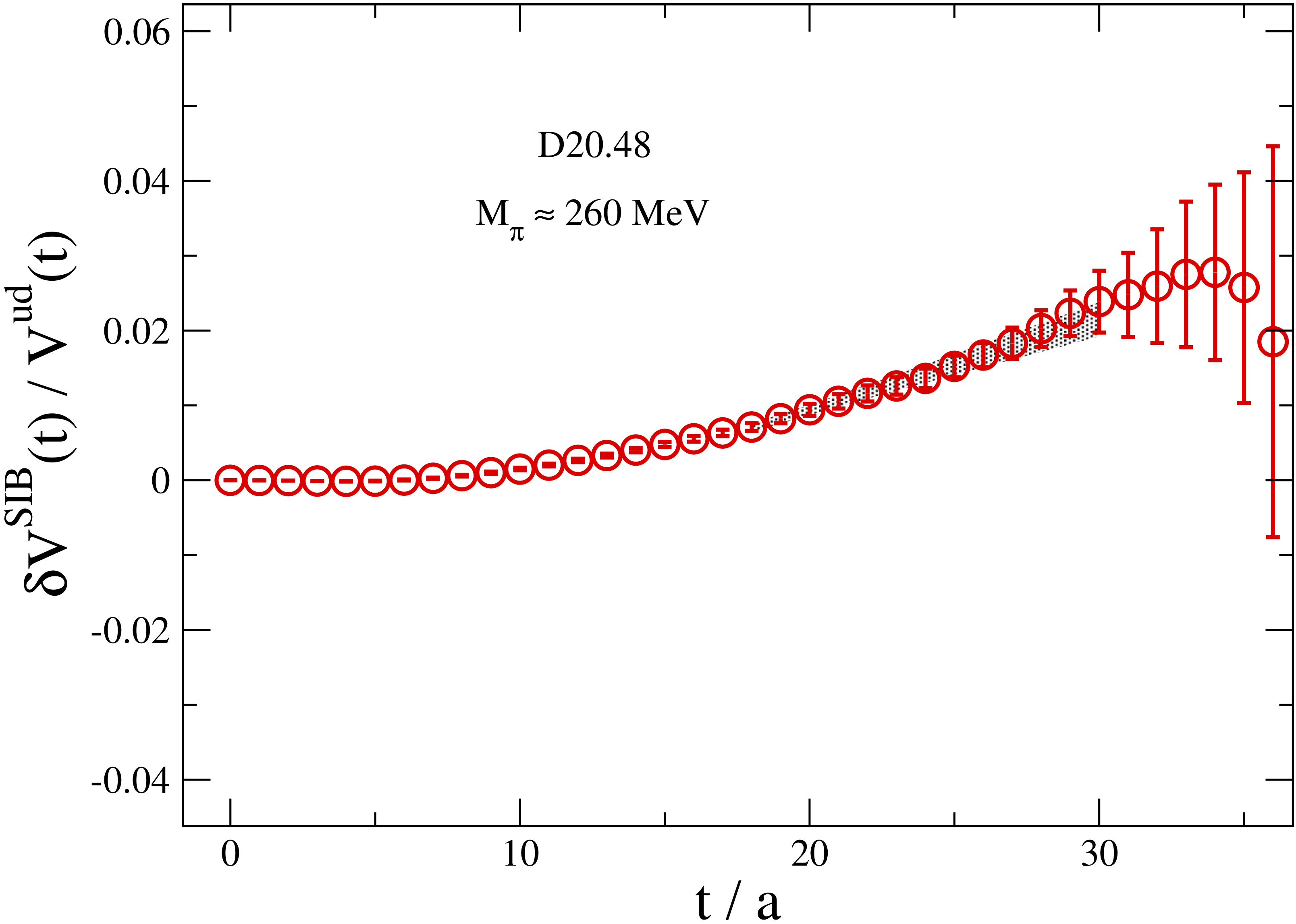}
\end{center}
\vspace{-0.50cm}
\caption{\it \small Ratios $\delta V^{QED} (t) / V^{ud} (t)$ (left panel) and $\delta V^{SIB} (t) / V^{ud} (t)$ (right panel) in the case of the gauge ensemble $D20.48$ versus the time distance $t$. The shaded areas correspond to the time interval where the ground-state is dominant (see Table~\ref{tab:GS_light}), together with the uncertainty (at $1 \sigma$ level) of a linear fit applied to the data.}
\label{fig:dVud_V}
\end{figure}

In Eq.~(\ref{eq:deltamu_cuts_II}) the quantities $\delta M_V^{ud}$ and $\delta Z_V^{ud}$ can be extracted respectively from the ``slope'' and the ``intercept'' of the ratio $\delta V^{ud}(t) / V^{ud}(t)$ at large time distances (see Refs.~\cite{deDivitiis:2011eh,deDivitiis:2013xla,Giusti:2017dmp,Giusti:2017jof}), namely
 \bea
      \frac{ \delta V^{ud}(t)}{V^{ud}(t)} ~ _{\overrightarrow{t >> a, (T-t) >> a}} ~ \frac{\delta Z_V^{ud}}{Z_V^{ud} } + 
                                                             \frac{\delta M_V^{ud}}{M_V^{ud}} f_{ud}(t)
      \label{eq:ratio}
 \eea
where
 \be
      f_{ud}(t) \equiv  M_V^{ud} \left( \frac{T}{2} - t \right) \frac{e^{- M_V^{ud} t} - e^{- M_V^{ud} (T-t)}}{e^{- M_V^{ud} t} + e^{- M_V^{ud} (T-t)}} - 
                               1 - M_V^{ud} \frac{T}{2} \approx - \left( 1 + M_V^{ud} t \right)
      \label{eq:fud}
 \ee
is almost a linear function of the Euclidean time $t$.
This procedure is shown in Fig.~\ref{fig:dVud_V} in the case of the gauge ensemble $D20.48$.

The time dependencies of the integrand functions $K_\mu(t) \, \delta V^{QED}(t)$ and $K_\mu(t) \, \delta V^{SIB}(t)$ are shown in Fig.~\ref{fig:damut} in the case of the ETMC gauge ensemble $B55.32$ (see Appendix~\ref{sec:appA}).
After summation over the time distance $t$, the SIB contribution dominates over the QED one.

\begin{figure}[htb!]
\begin{center}
\includegraphics[scale=0.370]{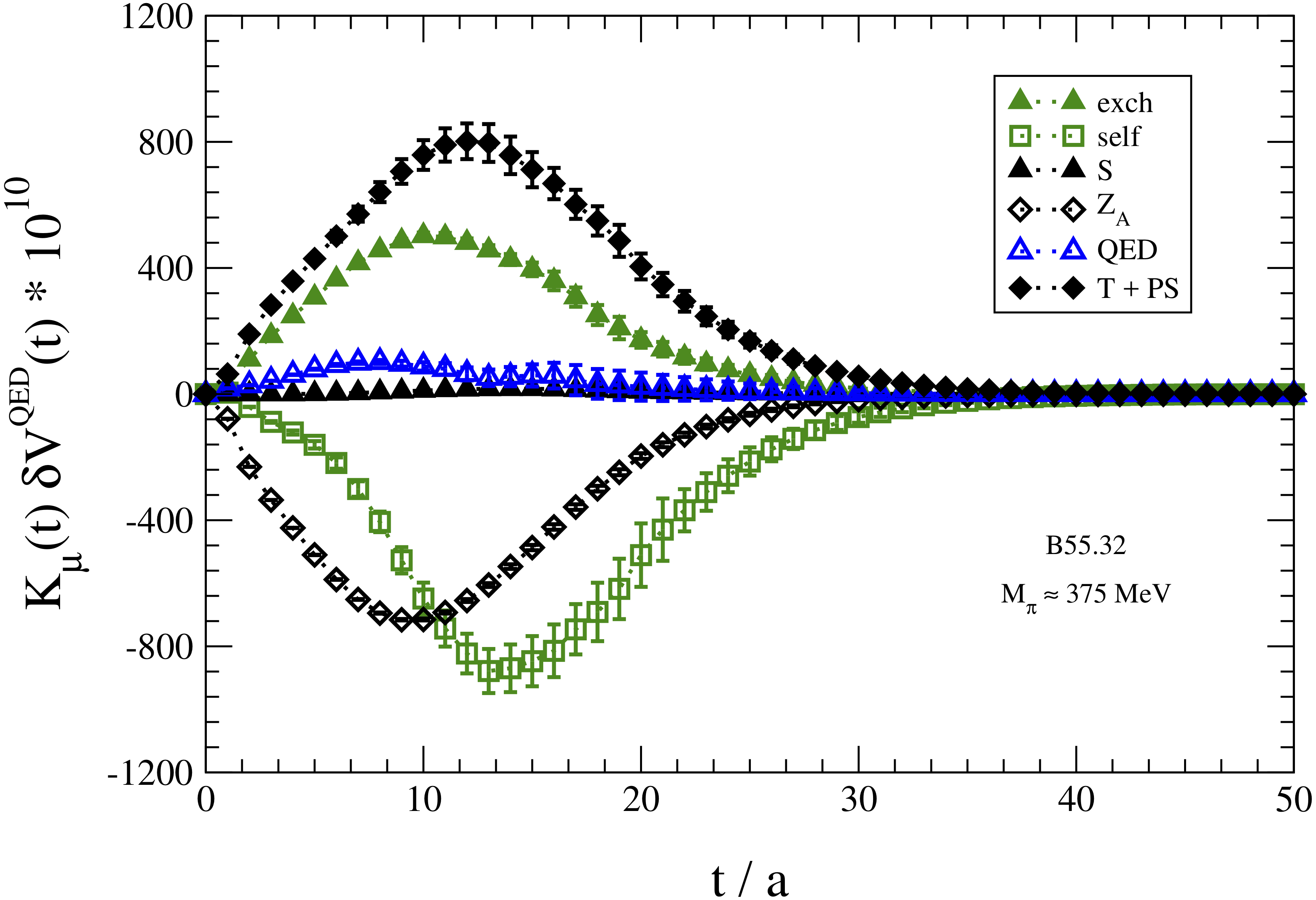} ~~ \includegraphics[scale=0.370]{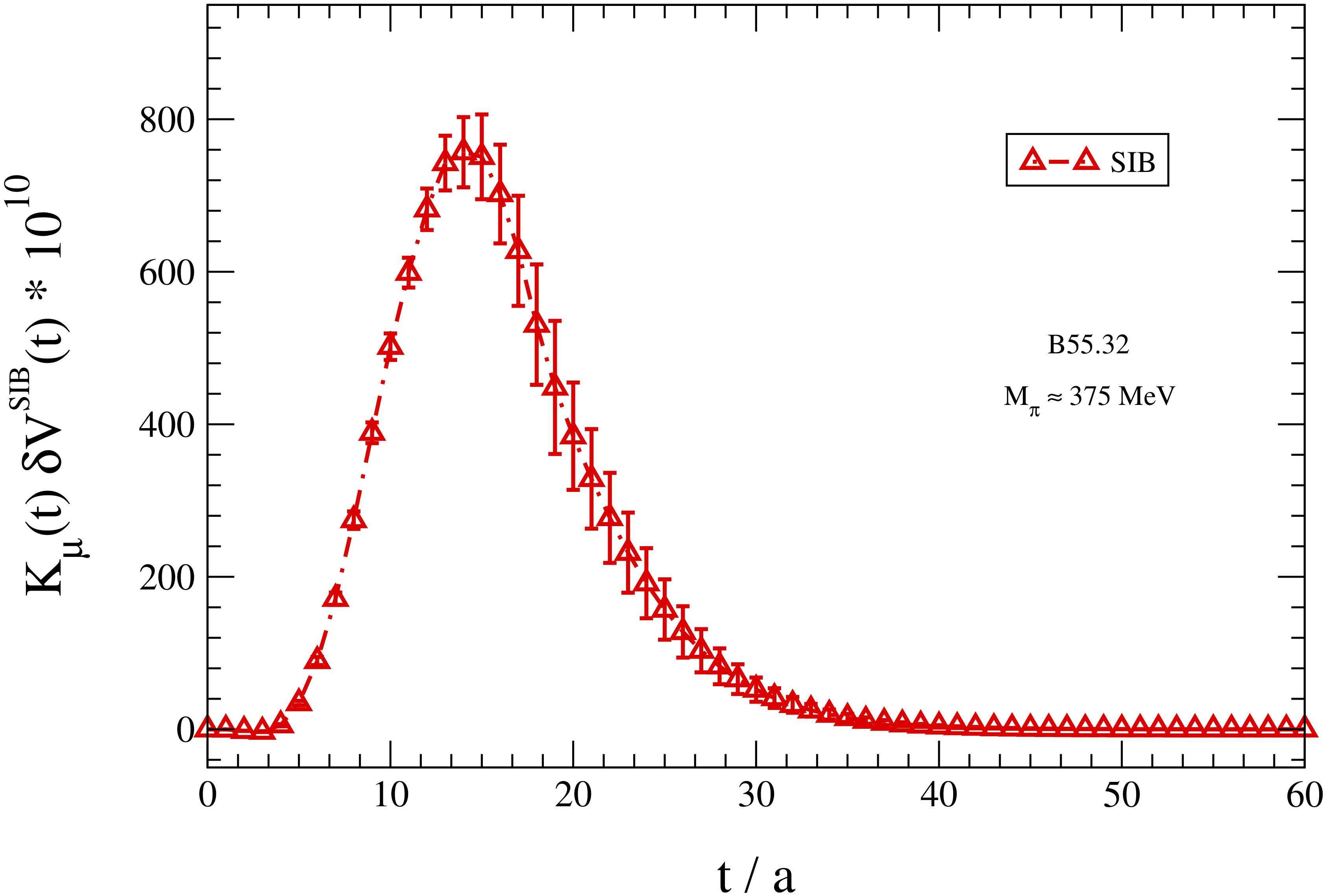}
\end{center}
\vspace{-0.55cm}
\caption{\it \small Time dependence of the integrand functions $K_\mu(t) \, \delta V^{QED}(t)$ (left panel) and $K_\mu(t) \, \delta V^{SIB}(t)$ (right panel) for the $u$- and $d$-quark contributions to the IB corrections $\delta a_\mu^{ud}$ [see Eq.~(\ref{eq:deltamu_cuts_I})] in the case of  the ETMC gauge ensemble $B55.32$.  The simulated pion mass is $M_\pi \simeq 375$ MeV and the lattice spacing is equal to $a \simeq 0.082$ fm. In the left panel the labels ``self'', ``exch'', ``T+PS'', ``S', ``${\cal Z}_A$'' indicate the QED contributions of the diagrams (\ref{fig:diagrams}a), (\ref{fig:diagrams}b), (\ref{fig:diagrams}c)+(\ref{fig:diagrams}d), (\ref{fig:diagrams}e) and the one generated by the QED corrections to the RC $Z_A$ of the local vector current [see Eq.~(\ref{eq:deltaV_ZA})].}
\label{fig:damut}
\end{figure} 

The results for the separate contributions $\delta a_\mu^{\rm HVP}(<)$ and $\delta a_\mu^{\rm HVP}(>)$, as well as their sum $\delta a_\mu^{\rm HVP}(ud)$, are obtained adopting four choices of\,~$T_{data}$, namely: $T_{data} = (t_{min} + 2a)$, $(t_{min} + t_{max}) / 2$, $(t_{max} - 2a)$ and $(T / 2 - 4a)$.
These results are collected in Table~\ref{tab:Tdata} for some of the ETMC gauge ensembles.
We find that the separation between $\delta a_\mu^{\rm HVP}(<)$ and $\delta a_\mu^{\rm HVP}(>)$ depends on the specific value of $T_{data}$, as it should be, but their sum $\delta a_\mu^{\rm HVP}(ud)$ is independent of the specific choice of the value of $T_{data}$ within the statistical uncertainties.
Note that for $T_{data} = t_{max} - 2a$ the contribution $\delta a_\mu^{\rm HVP}(>)$, which depends on the identification of the ground-state signal, is still a significant fraction of the total value $\delta a_\mu^{\rm HVP}(ud)$, as it was already observed in the case of the lowest-order term $a_\mu^{\rm HVP}(ud)$ in Ref.~\cite{Giusti:2018mdh}.
\begin{table}[hbt!]
\begin{center}
\centerline{ensemble A80.24} 
\vspace{0.25cm}
\begin{tabular}{||c||c|c|c|c||}
\hline
$T_{data}$ & $ (t_{min} + 2a) $ & $ (t_{min} + t_{max}) / 2 $ &  $ (t_{max} - 2a) $ &  $ (T / 2 - 4a) $ \\
\hline \hline
$\delta a_\mu^{\rm HVP}(<)$ & $0.83 ~~ (5)$ & $1.00 ~~ (4)$ & $1.13 ~~ (5)$ & $1.26 ~~ (7)$ \\
\hline
$\delta a_\mu^{\rm HVP}(>)$ & $0.61 ~ (12)$ & $0.43 ~ (10)$ & $0.30 ~ (8)$ & $0.18~~ (6)$ \\
\hline
$\delta a_\mu^{\rm HVP}(ud)$ & $1.44 ~ (12)$ & $1.43 ~ (12)$ & $1.43 ~ (12)$ & $1.44 ~ (12)$ \\
\hline \hline
\end{tabular}

\vspace{0.5cm}

\centerline{ensemble A50.32} 
\vspace{0.25cm}
\begin{tabular}{||c||c|c|c|c||}
\hline
$T_{data}$ & $ (t_{min} + 2a) $ & $ (t_{min} + t_{max}) / 2 $ &  $ (t_{max} - 2a) $ &  $ (T / 2 - 4a) $ \\
\hline \hline
$\delta a_\mu^{\rm HVP}(<)$ & $1.07 ~~ (8)$ & $1.45 ~ (14)$ & $1.65 ~ (20)$ & $1.62 ~ (28)$ \\
\hline
$\delta a_\mu^{\rm HVP}(>)$ & $0.73 ~ (26)$ & $0.39 ~ (19)$ & $0.20 ~ (12)$ & $0.02 ~~ (2)$ \\
\hline
$\delta a_\mu^{\rm HVP}(ud)$ & $1.80 ~ (30)$ & $1.84 ~ (29)$ & $1.85~ (29)$ & $1.64 ~ (29)$ \\
\hline \hline
\end{tabular}

\vspace{0.5cm}

\centerline{ensemble B55.32} 
\vspace{0.25cm}
\begin{tabular}{||c||c|c|c|c||}
\hline
$T_{data}$ & $ (t_{min} + 2a) $ & $ (t_{min} + t_{max}) / 2 $ &  $ (t_{max} - 2a) $ &  $ (T / 2 - 4a) $ \\
\hline \hline
$\delta a_\mu^{\rm HVP}(<)$ & $1.03 ~~ (4)$ & $1.31 ~~ (6)$ & $1.53 ~~ (9)$ & $1.70 ~ (19)$ \\
\hline
$\delta a_\mu^{\rm HVP}(>)$ & $0.64 ~ (18)$ & $0.40 ~ (15)$ & $0.20 ~ (10)$ & $0.03 ~~ (2)$ \\
\hline
$\delta a_\mu^{\rm HVP}(ud)$ & $1.67 ~ (20)$ & $1.71 ~ (20)$ & $1.73 ~ (19)$ & $1.73 ~ (21)$ \\
\hline \hline
\end{tabular}

\vspace{0.5cm}

\centerline{ensemble D30.48} 
\vspace{0.25cm}
\begin{tabular}{||c||c|c|c|c||}
\hline
$T_{data}$ & $ (t_{min} + 2a) $ & $ (t_{min} + t_{max}) / 2 $ &  $ (t_{max} - 2a) $ &  $ (T / 2 - 4a) $ \\
\hline \hline
$\delta a_\mu^{\rm HVP}(<)$ & $1.10 ~~ (7)$ & $1.55 ~ (12)$ & $2.05 ~ (18)$ & $2.42 ~ (63)$ \\
\hline
$\delta a_\mu^{\rm HVP}(>)$ & $2.01 ~ (30)$ & $1.51 ~ (25)$ & $0.99 ~ (18)$ & $0.09 ~~ (2)$ \\
\hline
$\delta a_\mu^{\rm HVP}(ud)$ & $3.11 ~ (35)$ & $3.06 ~ (34)$ & $3.04 ~ (34)$ & $2.51 ~ (63)$ \\
\hline \hline
\end{tabular}

\end{center}
\caption{\it \small Results for the contributions $\delta a_\mu^{\rm HVP}(<)$, $\delta a_\mu^{\rm HVP}(>)$ and their sum $\delta a_\mu^{\rm HVP}(ud)$, in units of $10^{-10}$, obtained adopting in Eqs.~(\ref{eq:deltamu_cuts_I}-\ref{eq:deltamu_cuts_II}) four different choices of $T_{data}$, namely: $T_{data} = (t_{min} + 2a)$, $(t_{min} + t_{max}) / 2$, $(t_{max} - 2a)$ and $(T / 2 - 4a)$ for the ETMC gauge ensembles $A80.24$, $A50.32$, $B55.32$ and $D30.48$ (see Table~\ref{tab:GS_light} for the values of $t_{min}$ and $t_{max}$, and Table~\ref{tab:simudetails} of Appendix~\ref{sec:appA} for the ETMC gauge ensembles).}
\label{tab:Tdata}
\end{table}

All four choices of $T_{data}$ are employed in the various branches of our bootstrap analysis.
The corresponding systematics is sub-dominant with respect to the other sources of uncertainties and it will not be given separately in the final error budget.

\begin{figure}[htb!]
\begin{center}
\includegraphics[scale=0.375]{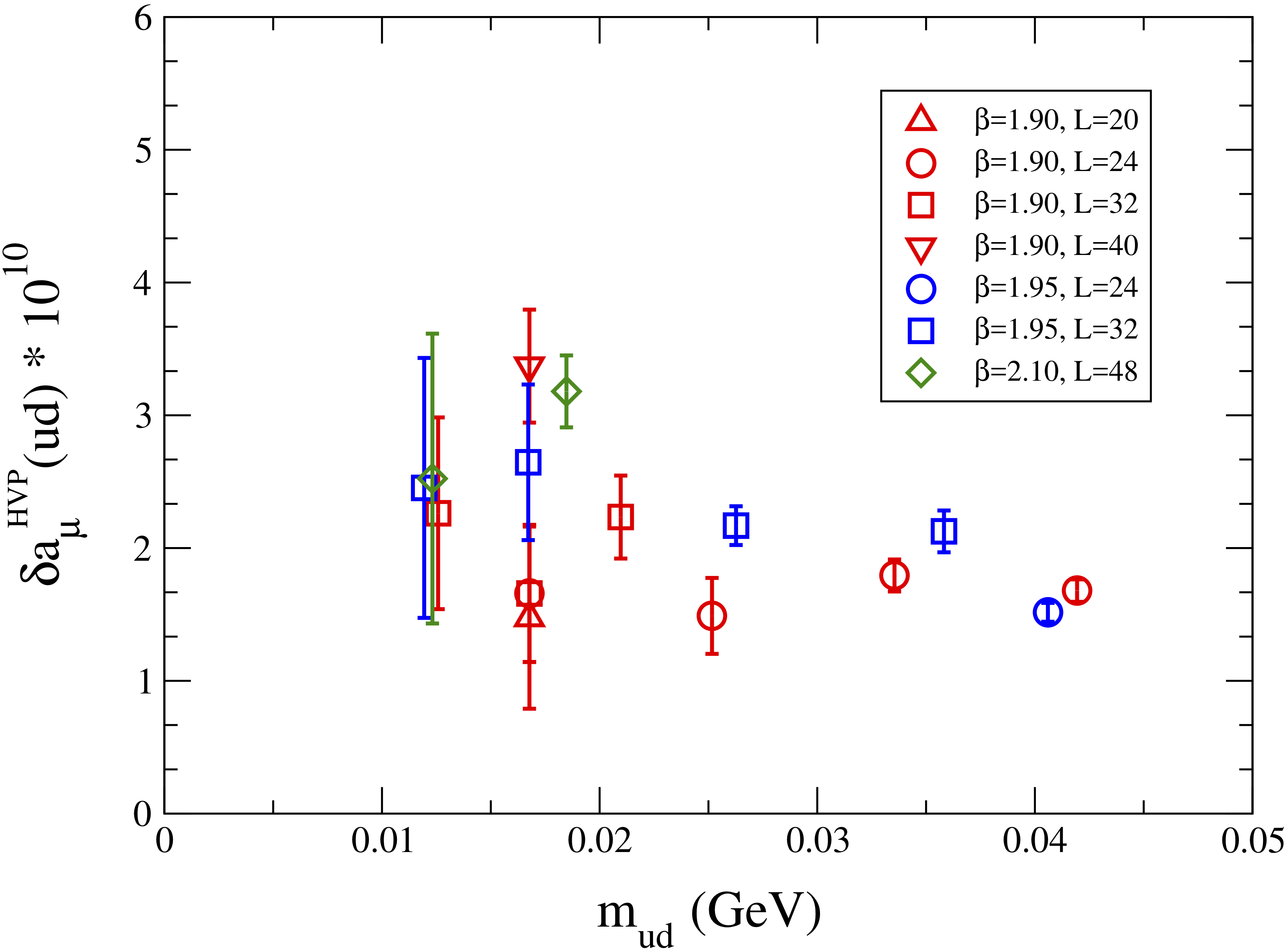} ~~ \includegraphics[scale=0.375]{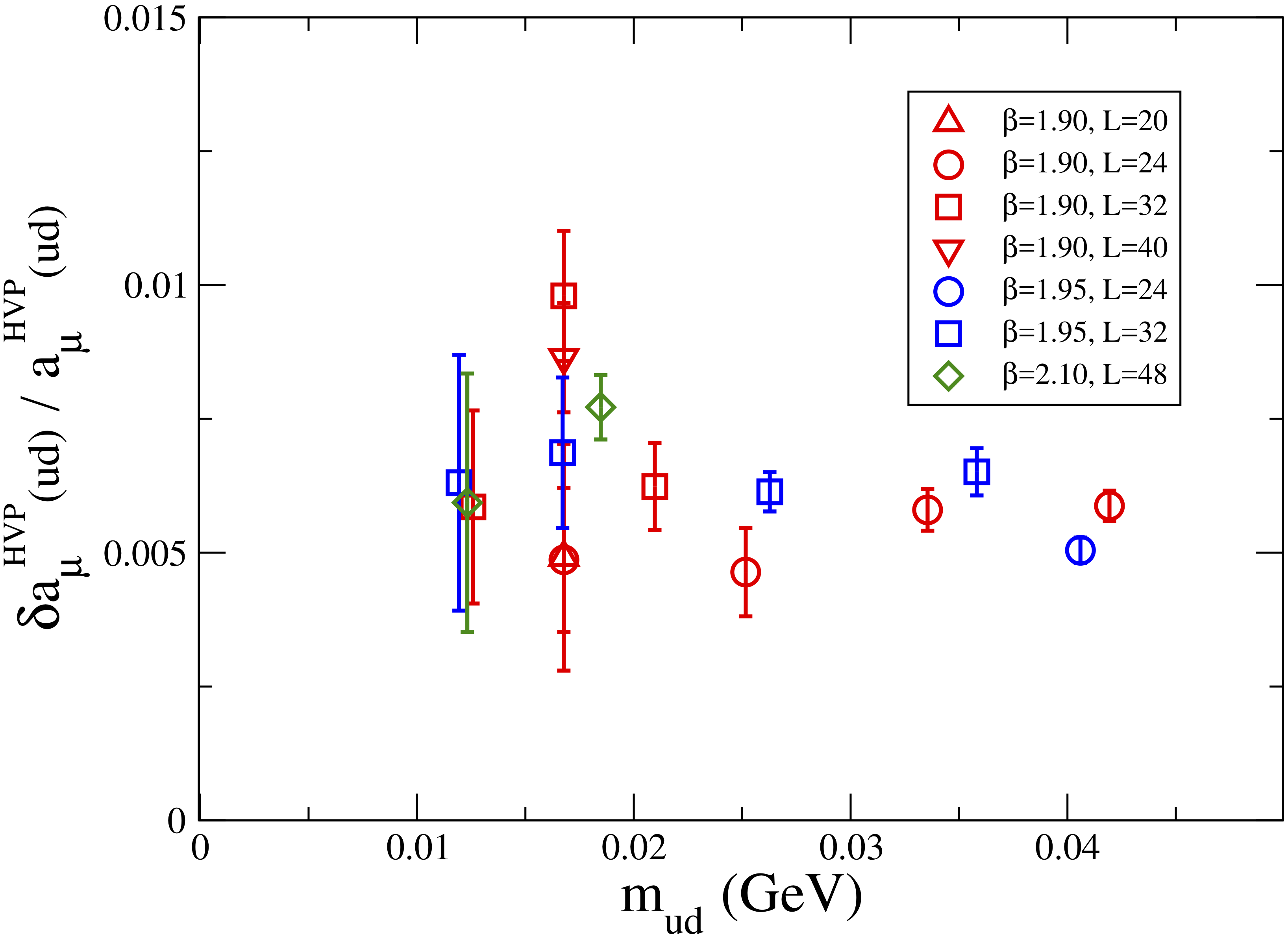}
\end{center}
\vspace{-0.75cm}
\caption{\it \small Results for $\delta a_\mu^{\rm HVP} (ud)$ (left panel) and the ratio $\delta a_\mu^{\rm HVP}(ud) / a_\mu^{\rm HVP}(ud)$ (right panel) versus the renormalized average $u/d$ mass $m_{ud}$ (in the $\overline{\rm MS}(2 ~ \rm GeV)$ scheme). Errors are the quadrature of the statistical uncertainties and of the error generated by the uncertainties of the input parameters of the quark mass analysis of Ref.~\cite{Carrasco:2014cwa} (see Appendix~\ref{sec:appA}).}
\label{fig:ratio_ud}
\end{figure}

We have considered also the ratio of the IB correction $\delta a_\mu^{\rm HVP}(ud)$ over the leading-order term $a_\mu^{\rm HVP} (ud)$, which was evaluated in Ref.~\cite{Giusti:2018mdh} for the same gauge ensembles.
The attractive feature of the ratio $\delta a_\mu^{\rm HVP}(ud) / a_\mu^{\rm HVP}(ud)$ is to be less sensitive to some of the systematic effects, in particular to the uncertainties of the scale setting. 
The data for $\delta a_\mu^{\rm HVP}(ud)$ and the ratio $\delta a_\mu^{\rm HVP}(ud) / a_\mu^{\rm HVP}(ud)$ are shown respectively in the left and right panels of Fig.~\ref{fig:ratio_ud}.
It can be seen that discretization effects play a minor role, while FVEs are more relevant.

For the separate QED and SIB contributions the FVEs differ qualitatively and quantitatively. 
In the case of the QED data a power-law behavior in terms of the inverse lattice size $1/ L$ is expected.
According to the general findings of Ref.~\cite{Lubicz:2016xro} the universal, structure-indepedent FVEs are expected to vanish, since they depend on the global charge of the meson states appearing in the spectral decomposition of the vector correlator, while the structure-dependent (SD) FVEs start at order ${\cal{O}}(1/L^2)$. 
Moreover, using the effective field theory approach of Ref.~\cite{Davoudi:2014qua} one may argue that in the case of mesons with vanishing charge radius (as the ones appearing in the vector correlator) the SD FVEs start at order ${\cal{O}}(1/L^3)$ (see also Ref.~\cite{Giusti:2017jof}). 
In the case of the SIB correlator (\ref{eq:deltaV_SIB}), since a fixed value $m_d - m_u = 2.38 \, (18)$ MeV~\cite{Giusti:2017dmp} is adopted for all gauge ensembles, an exponential dependence in terms of the quantity $M_\pi L$ is expected~\cite{Lin:2001ek}.

\begin{figure}[htb!]
\begin{center}
\includegraphics[scale=0.70]{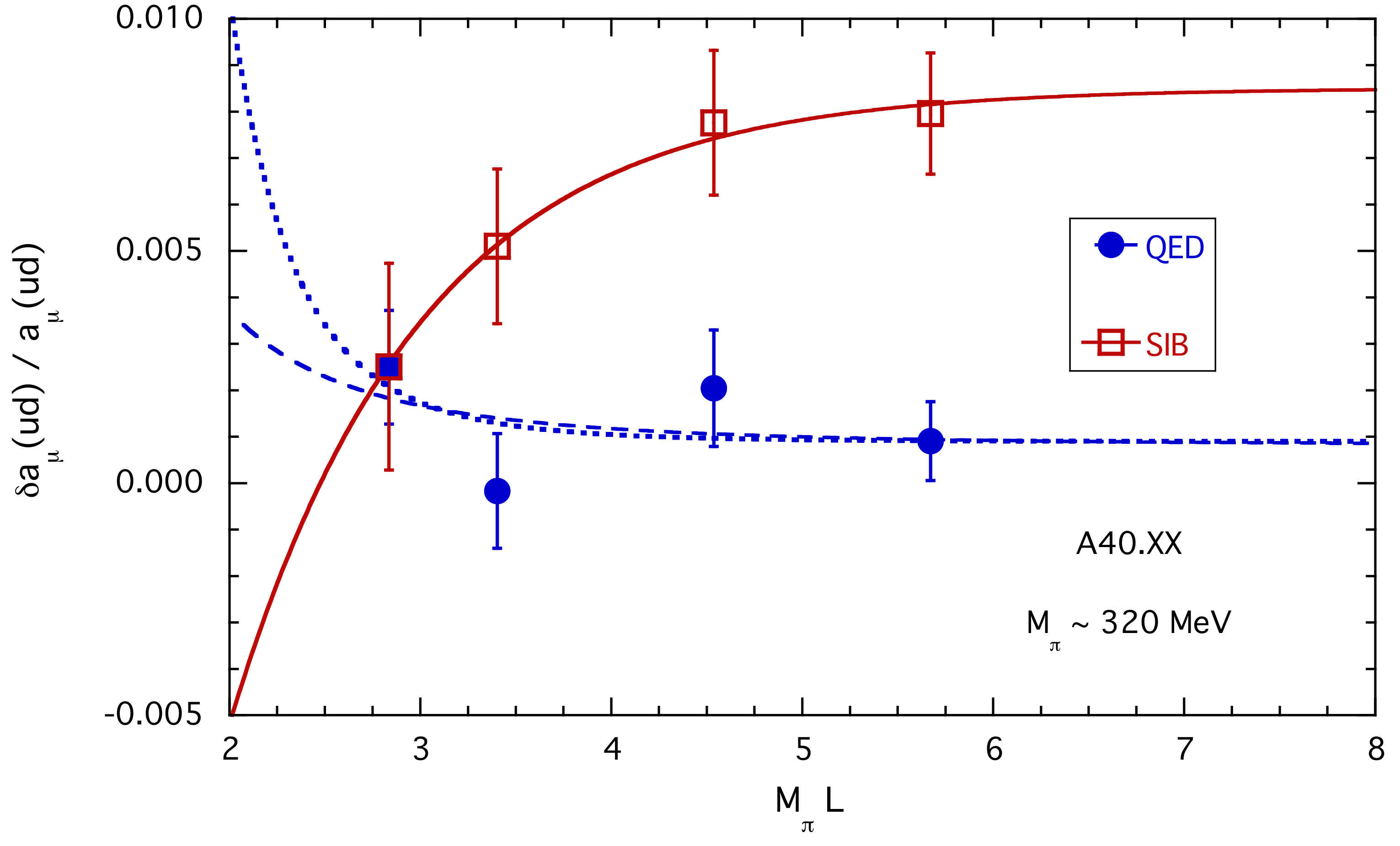}
\end{center}
\vspace{-0.75cm}
\caption{\it \small Results for the ratio $\delta a_\mu^{\rm HVP} (ud) / a_\mu^{\rm HVP} (ud)$ versus the quantity $M_\pi L$ in the case of the four ensembles A40.XX, which share common values of the light-quark mass and of the lattice spacing, but differ in the lattice size $L$. The empty (full) markers correspond to the SIB (QED) contribution. The solid line is a fit of the SIB data using the phenomenological Ansatz $A + B e^{-M_\pi L}$. The dashed and dotted lines correspond to a fitting function of the form $\overline{A} + \overline{B} / L^n$ with $n = 3$ (dashed line) and $n = 6$ (dotted line) both applied to the QED data.}
\label{fig:ratio_FVE}
\end{figure}

In Fig.~\ref{fig:ratio_FVE} the data for the QED and SIB contributions to the ratio $\delta a_\mu^{\rm HVP} (ud) / a_\mu^{\rm HVP} (ud)$ are shown in the case of the four ensembles A40.XX, which share common values of the light-quark mass and of the lattice spacing, but differ in the lattice size $L$. 
It can be seen that the theoretical expectations for the FVEs are consistent with the lattice data for both the QED and SIB contributions\footnote{We remind the reader that the lowest-order term $a_\mu^{\rm HVP} (ud)$ has nonnegligible FVEs, which are exponentially suppressed in terms of $M_\pi L$~\cite{Lin:2001ek} (see Fig.~9 of Ref.~\cite{Giusti:2018mdh}).}.

Since the SIB data dominate over the QED ones, the FVEs for the ratio $\delta a_\mu^{\rm HVP} (ud) / a_\mu^{\rm HVP} (ud)$ are expected to be mainly exponentially suppressed in $M_\pi L$.

For the combined extrapolations to the physical pion mass and to the continuum and infinite-volume limits we adopt the following fit ansatz:
\be
    \frac{\delta a_\mu^{\rm HVP} (ud)}{a_\mu^{\rm HVP} (ud)} = \delta_0 \left [ 1 + \delta_1 \, m_{ud} + \delta_{1l} \, m_{ud} \, \mbox{ln}(m_{ud}) + 
         \delta_2 \, m_{ud}^2 + D \, a^2 + \delta_{FVE} \right] ~ ,
\label{eq:fit_ansatz}
\ee
where the FVE term is estimated by using alternatively one of the fitting functions
\bea
\delta_{FVE} & = & F ~ e^{- \overline{M} L} ~ \qquad \qquad \qquad \qquad \mbox{or} ~ \nonumber \\[2mm]
\delta_{FVE} & = & \widehat{F}_n ~ \frac{\overline{M}^2}{16 \pi^2 f_0^2} ~ \frac{e^{-\overline{M} L}}{(\overline{M} L)^n} ~ \quad \qquad ({\rm n =} \frac{1}{2}, ~ 1, ~ \frac{3}{2}, ~ 2) ~ 
\eea
with $B_0$ and $f_0$ being the leading-order low-energy constants of ChPT and $\overline{M}^2 \equiv 2 B_0 m_{ud}$.
For the chiral extrapolation we consider either a quadratic ($\delta_{1l} = 0$ and $\delta_2 \neq 0$) or a logarithmic ($\delta_{1l} \neq 0$ and $\delta_2 = 0$) dependence.
Half of the difference of the corresponding results extrapolated to the physical pion mass is used to estimate the systematic uncertainty due to the chiral extrapolation.
Discretization effects are estimated by including $(D \neq 0)$ or excluding $(D = 0)$ the term proportional to $a^2$ in Eq.~(\ref{eq:fit_ansatz}).
The free parameters to be determined by the fitting procedure are $\delta_0$, $\delta_1$, $\delta_{1l}$ (or $\delta_2$), $D$ and $F$ (or $\widehat{F}_n$).

\begin{figure}[htb!]
\begin{center}
\includegraphics[scale=0.7]{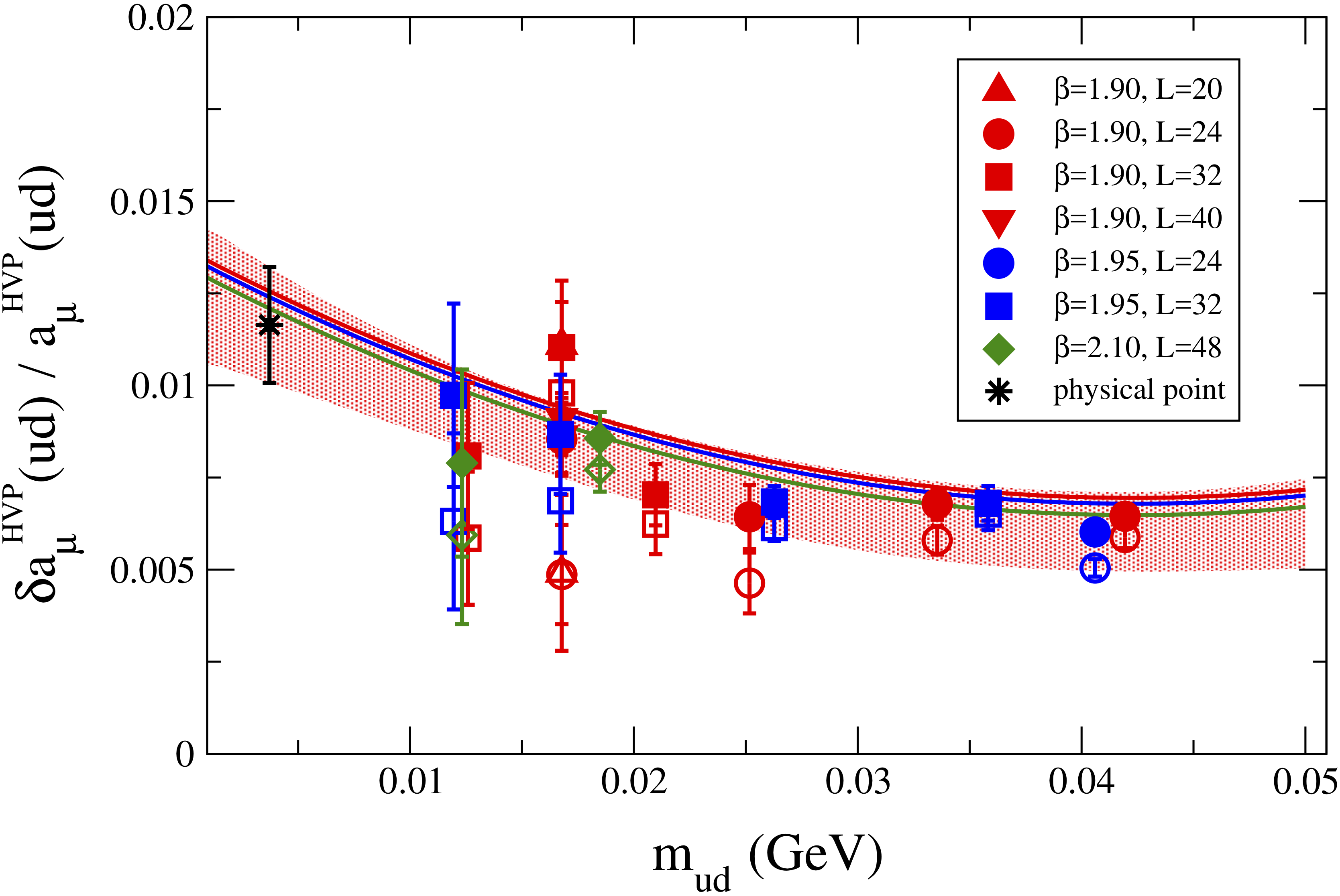}
\end{center}
\vspace{-0.75cm}
\caption{\it \small Results for the ratio $\delta a_\mu^{\rm HVP} (ud) / a_\mu^{\rm HVP} (ud)$ versus the renormalized average u/d mass $m_{ud}$ in the $\overline{\rm MS}(2 ~ \rm GeV)$ scheme. The empty markers correspond to the raw data, while the full ones represent the lattice data corrected by the FVEs obtained in the fitting procedure~(\ref{eq:fit_ansatz}) with $\delta_{1l} = 0$ and $\delta_2 \neq 0$. The solid lines correspond to the results of the combined fit~(\ref{eq:fit_ansatz}) obtained in the infinite-volume limit at each value of the lattice spacing. The black asterisk represents the value of the ratio $\delta a_\mu^{\rm HVP} (ud) / a_\mu^{\rm HVP} (ud)$ extrapolated to the physical pion mass, corresponding to $m^{phys}_{ud} (\overline{\rm MS}(2 ~ \rm GeV) = 3.70 ~ (17) ~ \rm MeV$ and to the continuum limit, while the red area indicates the corresponding uncertainty as a function of $m_{ud}$ at the level of one standard deviation. Errors are statistical only.}
\label{fig:ratio_ud_IB_RM123}
\end{figure}

In our combined fit (\ref{eq:fit_ansatz}) the values of the free parameters are determined by a $\chi^2$-minimization procedure adopting an uncorrelated $\chi^2$.
The uncertainties on the fitting parameters do not depend on the $\chi^2$-value, because they are obtained by using the bootstrap samplings of Ref.~\cite{Carrasco:2014cwa}.
This guarantees that all the correlations among the lattice data points and among the fitting parameters are properly taken into account.
The quality of our fitting procedure is illustrated in Fig.~\ref{fig:ratio_ud_IB_RM123}.

At the physical pion mass and in the continuum and infinite-volume limits we get
\be
    \frac{\delta a_\mu^{\rm HVP} (ud)}{a_\mu^{\rm HVP} (ud)} = 0.0115 ~ (18)_{stat+fit} \,  (21)_{input} \,  (20)_{chir} \, (19)_{\rm FVE}\, (9)_{a^2} \, [41] ~ ,
    \label{eq:ratio_IB}
\ee
where the errors come in the order from (statistics + fitting procedure), input parameters of the eight branches of the quark mass analysis of Ref.~\cite{Carrasco:2014cwa}, chiral extrapolation, finite-volume and discretization effects.
In Eq.~(\ref{eq:ratio_IB}) the uncertainty in the square brackets corresponds to the sum in quadrature of the statistical and systematic errors.

Using the leading-order result $a^{\rm HVP}_\mu (ud) = 619.0 ~ (17.8) \cdot 10^{-10}$ from Ref.~\cite{Giusti:2018mdh}, we obtain our determination of the leading-order IB corrections to $a^{\rm HVP}_\mu (ud)$, namely
\bea
    \delta a_\mu^{\rm HVP} (ud) & = & 7.1 ~ (1.1)_{stat+fit}\, (1.3)_{input}\, (1.2)_{chir}\, (1.2)_{\rm FVE}\, (0.6)_{a^2}\, \cdot 10^{-10} \nonumber \\
                                                  & = & 7.1 ~ (2.5) \cdot 10^{-10} ~ ,
    \label{eq:delta_ud}
\eea
which comes (within the GRS prescription) from the sum of the QED contribution
\be
    \left [ \delta a^{\rm HVP}_\mu(ud) \right ]^{(QED)} = 1.1 ~ (1.0) \cdot 10^{-10} ~ 
    \label{eq:delta_ud_QED}
\ee
and of the SIB one
\be
    \left [ \delta a^{\rm HVP}_\mu(ud) \right ]^{(SIB)} = 6.0 ~ (2.3) \cdot 10^{-10} ~ .
    \label{eq:delta_ud_SIB}
\ee
The above results show that the IB correction~(\ref{eq:delta_ud}) is dominated by the strong $SU(2)$-breaking term, which corresponds roughly to $\approx 85 \%$ of $\delta a^{\rm HVP}_\mu (ud)$.

Our determination~(\ref{eq:delta_ud}), obtained with $N_f = 2 + 1 + 1$ dynamical flavors of sea quarks, agrees within the errors with and is more precise than both the phenomenological estimate $\delta a^{\rm HVP}_\mu (ud) = 7.8 ~ (5.1) \cdot 10^{-10}$, obtained by the BMW collaboration~\cite{Borsanyi:2017zdw} using results of the dispersive analysis of $e^+ e^-$ data~\cite{Jegerlehner:2017lbd}, and the lattice determination $\delta a^{\rm HVP}_\mu (ud) = 9.5 ~ (10.2) \cdot 10^{-10}$, obtained by the RBC/UKQCD collaboration~\cite{Blum:2018mom} at $N_f = 2 + 1$, which includes also one disconnected QED diagram.
Recently, adopting $N_f = 1 + 1 + 1 + 1$ simulations, the FNAL/HPQCD/MILC collaboration has found for the SIB contribution the value $\left[ \delta a^{\rm HVP}_\mu(ud) \right]^{(SIB)} = 9.0 ~ (4.5) \cdot 10^{-10}$~\cite{Chakraborty:2017tqp}.

Thanks to the recent nonperturbative evaluation of QCD+QED effects on the RCs of bilinear operators performed in Refs.~\cite{DiCarlo:2019thl} we can update the determinations of the strange $\delta a_\mu^{\rm HVP}(s)$ and charm $\delta a_\mu^{\rm HVP}(c)$ contributions to the IB effects made in Ref.~\cite{Giusti:2017jof}.
We get
\bea
    \label{eq:delta_s}
    \delta a_\mu^{\rm HVP}(s) & = & -0.0053 ~ (30)_{stat+fit}\, (13)_{input}\, (2)_{chir}\, (2)_{\rm FVE}\, (1)_{a^2} \cdot 10^{-10} \nonumber \\
                                               & = & -0.0053 ~ (33) \cdot 10^{-10} ~ , \\[2mm]
    \label{eq:delta_c}
    \delta a_\mu^{\rm HVP}(c) & = & 0.0182 ~ (35)_{stat+fit}\, (5)_{input}\, (1)_{chir}\, (3)_{\rm FVE}\, (1)_{a^2} \cdot 10^{-10} \nonumber \\
                                               & = & 0.0182 ~ (36) \cdot 10^{-10} ~ 
\eea
to be compared with $\delta a_\mu^{\rm HVP}(s) = -0.018 ~ (11) \cdot 10^{-10}$ and $\delta a_\mu^{\rm HVP}(c) = -0.030 ~ (13) \cdot 10^{-10}$ given in Ref.~\cite{Giusti:2017jof}.
The updated results confirm that the em corrections $\delta a_\mu^{\rm HVP} (s)$ and $\delta a_\mu^{\rm HVP} (c)$ are negligible with respect to the current uncertainties of the corresponding lowest-order terms $a_\mu^{\rm HVP}(s) = 53.1 ~ (2.5) \cdot 10^{-10}$ and $a_\mu^{\rm HVP}(c) = 14.75 ~ (0.56) \cdot 10^{-10}$~\cite{Giusti:2017jof}.
Recently~\cite{Blum:2018mom} in the case of the strange contribution the RBC/UKQCD collaboration has found the result $\delta a_\mu^{\rm HVP} (s) = -0.0149~(32) \cdot 10^{-10}$, which deviates from our finding (\ref{eq:delta_s}) by $\approx 2$ standard deviations.

The sum of our three results (\ref{eq:delta_ud}), (\ref{eq:delta_s}) and (\ref{eq:delta_c}) yields the contribution of quark-connected diagrams to $\delta a_\mu^{\rm HVP}$ within the qQED approximation, namely $\delta a_\mu^{\rm HVP}(udsc)|_{conn} = 7.1 ~ (2.6) \cdot 10^{-10}$.
Recently, in Ref.~\cite{Blum:2018mom} one QED disconnected diagram has been calculated in the case of the $u$- and $d$-quark contribution and found to be of the same order of the corresponding QED connected term. 
Thus, we estimate that the uncertainty related to the qQED approximation and to the neglect of quark-disconnected diagrams is approximately equal to  our QED contribution (\ref{eq:delta_ud_QED}), obtaining 
\be
    \label{eq:delta_udsc}
    \delta a_\mu^{\rm HVP}(udsc) = 7.1 ~ (2.6) ~ (1.2)_{qQED+disc} \cdot 10^{-10} = 7.1 ~ (2.9) \cdot 10^{-10} ~ ,
\ee
which represents the most accurate determination of the IB corrections to $a_\mu^{\rm HVP}$ to date.

Using the recent ETMC determinations of the lowest-order contributions of light, strange and charm quarks, $a^{\rm HVP}_\mu (ud) = 619.0 ~ (17.8) \cdot 10^{-10}$, $a^{\rm HVP}_\mu (s) = 53.1 ~ (2.5) \cdot 10^{-10}$ and $a^{\rm HVP}_\mu (c) = 14.75 ~ (0.56) \cdot 10^{-10}$~\cite{Giusti:2017jof,Giusti:2018mdh}, and an estimate of the lowest-order quark-disconnected diagrams, $a^{\rm HVP}_\mu (disc) = -12 ~ (4) \cdot 10^{-10}$, obtained using the results of Refs.~\cite{Borsanyi:2017zdw} and~\cite{Blum:2018mom}, our finding (\ref{eq:delta_udsc}) for the IB corrections leads to an HVP contribution to the muon ($g - 2$) equal to 
\be
    a_\mu^{\rm HVP} = 682 ~ (19) \cdot 10^{-10} ~ ,
\ee
which agrees within the errors with the recent determinations based on dispersive analyses of the experimental cross section data for $e^+ e^-$ annihilation into hadrons (see Ref.~\cite{Keshavarzi:2018mgv} and references therein).

\section{Conclusions}
\label{sec:conclusions}

We have presented a lattice calculation of the isospin-breaking corrections to the HVP contribution of light quarks to the anomalous magnetic moment of the muon at order ${\cal{O}}[\alpha^2_{em} (m_d - m_u) / \Lambda_{QCD}]$ in the light-quark mass difference and ${\cal{O}}(\alpha^3_{em})$ in the em coupling.
We have employed the gauge configurations generated by ETMC with $N_f = 2+1+1$ dynamical quarks at three values of the lattice spacing $(a \simeq 0.062 - 0.089~\mbox{fm})$ with pion masses in the range $M_\pi \simeq 210 - 450~\mbox{MeV}$ and with strange and charm quark masses tuned at their physical values determined in Ref.~\cite{Carrasco:2014cwa}.

The calculation of the IB corrections has been carried out adopting the RM123 approach of Refs.~\cite{deDivitiis:2011eh,deDivitiis:2013xla}, which is based on the expansion of the lattice path-integral in powers of the {\it small} parameters $(m_d - m_u) / \Lambda_{QCD}$ and $\alpha_{em}$, which are both of the order of ${\cal{O}}(1\%)$.

In this work we have taken into account only connected diagrams in which each quark flavor contributes separately.
The leading-order em contributions to the renormalization constant of the local version of the lattice vector current, adopted in this work, have been evaluated using a recent nonperturbative calculation performed within the $RI^\prime$-MOM scheme in Refs.~\cite{DiCarlo:2019thl} .
Thanks to that we have updated also the determinations of the strange $\delta a_\mu^{\rm HVP}(s)$ and charm $\delta a_\mu^{\rm HVP}(c)$ IB contributions made in Ref.~\cite{Giusti:2017jof}, obtaining a drastic improvement of the uncertainties.

Within the qQED approximation and neglecting quark-disconnected diagrams the main results of the present study are:
\bea
    \label{eq:delta_ud_final}
    \delta a_\mu^{\rm HVP}(ud) & = & 7.1 ~ (2.5) \cdot 10^{-10} ~ , \\[2mm]
    \label{eq:delta_s_final}
    \delta a_\mu^{\rm HVP}(s) & = & -0.0053 ~ (33) \cdot 10^{-10} ~ , \\[2mm]
    \label{eq:delta_c_final}
    \delta a_\mu^{\rm HVP}(c) & = & 0.0182 ~ (36) \cdot 10^{-10} ~ , 
\eea

Summing up the three contributions (\ref{eq:delta_ud_final})-(\ref{eq:delta_c_final}) and adding a further $\approx 15 \%$ uncertainty related to the qQED approximation and to the neglect of quark-disconnected diagrams, we get
\be
    \label{eq:delta_udsc_final}
    \delta a_\mu^{\rm HVP}(udsc) = 7.1 ~ (2.6) ~ (1.2)_{qQED+disc} \cdot 10^{-10} = 7.1 ~ (2.9) \cdot 10^{-10} ~ ,
\ee
which represents the most accurate determination of the IB corrections to $a_\mu^{\rm HVP}$ to date.

New QCD simulations with $N_f = 2 + 1 + 1$ dynamical quarks close to the physical pion point~\cite{Alexandrou:2018egz} and the evaluation of quark-disconnected diagrams are in progress.

\section*{Acknowledgments}
We gratefully acknowledge the CPU time provided by CINECA under the initiative INFN-LQCD123 on the Marconi KNL and SKL systems at CINECA (Italy).
We thank S.~Bacchio and B.~Kostrezwa for their help in setting up the interface and the parameters for the DD$\alpha$AMG library~\cite{Alexandrou:2016izb} used to evaluate the quark propagators. We thank B.~Kostrezwa for his help in the HMC simulations used to produce the A40.40 gauge ensemble with the tmLQCD software package~\cite{Jansen:2009xp,Abdel-Rehim:2013wba,Deuzeman:2013xaa}.
G.~M., V.L.~and S.S. thank MIUR (Italy) for partial support under Contract No.~PRIN 2015P5SBHT. G.~M.~thanks the partial support from ERC Ideas Advanced Grant No.~267985 ``DaMeSyFla''.

\appendix

\section{Simulation details}
\label{sec:appA}

The ETMC gauge ensembles used in this work are the same adopted in Ref.~\cite{Carrasco:2014cwa} to determine the up-, down-, strange- and charm-quark masses in isospin symmetric QCD.
We employ the Iwasaki action~\cite{Iwasaki:1985we} for gluons and the Wilson Twisted Mass Action~\cite{Frezzotti:2000nk,Frezzotti:2003ni,Frezzotti:2003xj} for sea quarks.
Working at maximal twist our setup guarantees an automatic ${\cal O} (a)$-improvement~\cite{Frezzotti:2003ni,Frezzotti:2004wz}.

We consider three values of the inverse bare lattice coupling $\beta$ and different lattice volumes, as shown in Table~\ref{tab:simudetails}, where the number of configurations analyzed $(N_{cfg})$ corresponds to a separation of 20 trajectories.
For earlier investigations of finite volume effects (FVEs) the ETMC had produced three dedicated ensembles, A40.20, A40.24 and A40.32, which share the same light-quark mass and lattice spacing and differ only in the lattice size $L$.
To improve such an investigation a further gauge ensemble, A40.40, has been generated at a larger value of the lattice size $L$.

At each lattice spacing, different values of the sea-light-quark masses are considered.
The valence- and sea-light-quark masses are always taken to be degenerate.
The values of the lattice spacing in isosymmetric QCD are: $a = 0.0885 ~ (36), ~ 0.0815 ~ (30), ~ 0.0619 ~ (18) ~ \mbox{fm}$ at $\beta = 1.90, ~ 1.95$ and $2.10$, respectively.

\begin{table}[hbt!]
{\small
\begin{center}
\begin{tabular}{||c|c|c||c|c|c|c|c|c||c|c|c||}
\hline
ensemble & $\beta$ & $V / a^4$ &$a\mu_{ud}$&$a\mu_\sigma$&$a\mu_\delta$&$N_{cfg}$& $M_\pi$ & $M_\pi L$  \\
\hline \hline
$A40.40$ & $1.90$ & $40^{3}\times 80$ &$0.0040$ &$0.15$ &$0.19$ & $100$ & 317 (12) & 5.7 \\
\cline{1-1} \cline{3-4} \cline{7-9}
$A30.32$ & & $32^{3}\times 64$ &$0.0030$ & & & $150$ & 275 (10) & 3.9 \\
$A40.32$ & & & $0.0040$ & & & $100$ & 316 (12) & 4.5  \\
$A50.32$ & & & $0.0050$ & & & $150$ & 350 (13) & 5.0  \\
\cline{1-1} \cline{3-4} \cline{7-9}
$A40.24$ & & $24^{3}\times 48 $ & $0.0040$ & & & $150$ & 322 (13) & 3.5 \\
$A60.24$ & & & $0.0060$ & & & $150$ & 386 (15) & 4.2 \\
$A80.24$ & & & $0.0080$ & & & $150$ & 442 (17) & 4.8 \\
$A100.24$ &  & & $0.0100$ & & & $150$ & 495 (19) & 5.3 \\
\cline{1-1} \cline{3-4} \cline{7-9}
$A40.20$ & & $20^{3}\times 48 $ & $0.0040$ & & & $150$ & 330 (13) & 3.0 \\
\hline \hline
$B25.32$ & $1.95$ & $32^{3}\times 64$ &$0.0025$&$0.135$ &$0.170$& $150$ & 259 (9) & 3.4 \\
$B35.32$ & & & $0.0035$ & & & $150$ & 302 (10) & 4.0 \\
$B55.32$ & & & $0.0055$ & & & $150$ & 375 (13) & 5.0 \\
$B75.32$ &  & & $0.0075$ & & & $~80$ & 436 (15) & 5.8 \\
\cline{1-1} \cline{3-4} \cline{7-9}
$B85.24$ & & $24^{3}\times 48 $ & $0.0085$ & & & $150$ & 468 (16) & 4.6 \\
\hline \hline
$D15.48$ & $2.10$ & $48^{3}\times 96$ &$0.0015$&$0.1200$ &$0.1385 $& $100$& 223 (6) & 3.4 \\ 
$D20.48$ & & & $0.0020$ & & & $100$ & 256 (7) & 3.0 \\
$D30.48$ & & & $0.0030$ & & & $100$ & 312 (8) & 4.7 \\
 \hline   
\end{tabular}
\end{center}
}
\vspace{-0.25cm}
\caption{\it \small Values of the simulated-quark bare masses (in lattice units), of the pion mass (in units of MeV) and of the product $M_\pi L$ for the $16$ ETMC gauge ensembles with $N_f = 2+1+1$ dynamical quarks used in this contribution (see Ref.~\cite{Carrasco:2014cwa}) and for the gauge ensemble, A40.40 added to improve the investigation of FVEs. The bare twisted masses $\mu_\sigma$ and $\mu_\delta$ describe the strange and charm sea doublet according to Ref.~\cite{Frezzotti:2003xj}. The central values and errors of the pion mass  are evaluated using the bootstrap events of the eight branches of the analysis of Ref.~\cite{Carrasco:2014cwa}. The valence quarks in the pion are regularized with opposite values of the Wilson $r$-parameter in order to guarantee that discretization effects on the pion mass are of order ${\cal{O}}(a^2 \mu_{ud} ~ \Lambda_{QCD})$.\hspace*{\fill}}
\label{tab:simudetails}
\end{table}

We made use of the bootstrap samplings elaborated for the input parameters of the quark mass analysis of Ref.~\cite{Carrasco:2014cwa}.
There, eight branches of the analysis were adopted differing in:
\begin{itemize}
\item the continuum extrapolation adopting for the scale parameter either the Sommer parameter $r_0$ or the mass of a fictitious pseudoscalar meson made up of strange(charm)-like quarks;
\item the chiral extrapolation performed with fitting functions chosen to be either a polynomial expansion or a Chiral Perturbation Theory (ChPT) Ansatz in the light-quark mass;
\item the choice between the methods M1 and M2, which differ by ${\cal{O}}(a^2)$ effects, used to determine the mass RC $Z_m = 1 / Z_P$ in the RI$^\prime$-MOM scheme.
\end{itemize}

Throughout this work the renormalized average $u/d$ quark mass $m_{ud}$ is given in the $\overline{\rm MS}$ scheme at a renormalization scale equal to $2$ GeV. 
We recall that, in the GRS prescription we have chosen, the renormalized average u/d quark mass in isosymmetric QCD $m_{ud}^{(0)}$ coincide with the one in QCD+QED, i.e.~$m_{ud} = m_{ud}^{(0)}$, in the $\overline{\rm MS}(2 \, {\rm GeV})$ scheme.
At the physical pion mass ($M_\pi^{phys} = M_{\pi^0} \simeq 135 \, \mbox{MeV}$) the value $m_{ud}^{phys} = 3.70 \, (17) \, \mbox{MeV}$ was determined in Ref.~\cite{Carrasco:2014cwa}, using the PDG value of the pion decay constant~\cite{PDG} for fixing the lattice scale.

The statistical accuracy of the meson correlator is based on the use of the so-called ``one-end'' stochastic method~\cite{McNeile:2006bz}, which includes spatial stochastic sources at a single time slice chosen randomly.
In the case of the light-quark contribution we have used 160 stochastic sources (diagonal in the spin variable and dense in the color one) for each gauge configuration.

Finally, the values evaluated in Ref.~\cite{DiCarlo:2019thl} for the coefficients $Z_m^{fact}$ [see Eq.~(\ref{eq:Zm})] and $Z_A^{fact}$ [see Eq.~(\ref{eq:ZA})] are collected in Table~\ref{tab:Zfact}.

\begin{table}[htb!]
\begin{center}
\begin{tabular}{||c||c|c||c|c||}
\hline
$\beta$ & $Z_m^{fact} ~ \rm{(M1)}$ & $Z_A^{fact} ~ \rm{(M1)}$ & $Z_m^{fact} ~ \rm{(M2)}$ & $Z_A^{fact} ~\rm{(M2)}$ \\
\hline \hline
$ ~ 1.90 ~ $ & ~ 1.629 \, (41) ~ & ~ 0.859 \, (15) ~ & ~ 1.637 \, (14) ~ & ~ 0.990 \, (9) ~ \\ 
\hline
$ ~ 1.95 ~ $ & ~ 1.514 \, (33) ~ & ~ 0.873 \, (13) ~ & ~ 1.585 \, (12) ~ & ~ 0.980 \, (8) ~ \\ 
\hline
$ ~ 2.10 ~ $ & ~ 1.459 \, (17) ~ & ~ 0.909 \,~ (6) ~ & ~ 1.462 \,~ (6) ~ & ~ 0.958 \, (3) ~ \\ 
\hline  
\end{tabular}
\end{center}
\vspace{-0.25cm}
\caption{\it \small Values adopted for the coefficients $Z_m^{fact}$ [see Eq.~(\ref{eq:Zm})] and $Z_A^{fact}$ [see Eq.~(\ref{eq:ZA})] evaluated in Ref.~\cite{DiCarlo:2019thl} for the M1 and M2 renormalization methods (see Ref.~\cite{Carrasco:2014cwa}) at the three values of $\beta$. In Ref.~\cite{Giusti:2017jof} a common value $Z_A^{fact} = 0.9 \, (1)$, estimated through the axial Ward-Takahashi identity derived in the presence of QED effects, was adopted at all values of $\beta$.\hspace*{\fill}}
\label{tab:Zfact}
\end{table}

\end{document}